\documentstyle[prd,aps,eqsecnum]{revtex}

\def\be{\begin{equation}}
\def\ee{\end{equation}}
\def\bea{\begin{eqnarray}}
\def\nn{\nonumber}
\def\eea{\end{eqnarray}}

\def\deriv{\hspace{-1.5ex}\mbox{ }^{\mbox{ }
  ^{\overline{\mbox{\hspace{1.5ex}}}}}
  \hspace{-1.89ex}\bigtriangledown}
\def\bbderiv{\hspace{-1.5ex}\mbox{ }^{\mbox{ }
  ^{\overline{\mbox{\hspace{1.5ex}}}}}
  \hspace{-2.5ex}\bigtriangledown}
\def\a{\alpha}
\def\g{\gamma}
\def\d{\delta}
\def\s{\sigma}

\begin{document}

\draft

\title{Stochastic semiclassical gravity}
\author{Rosario Mart\'{\i}n and Enric Verdaguer\footnote{Institut de 
                              F\'{\i}sica d'Altes Energies (IFAE)}}
\address{Departament de F\'{\i}sica Fonamental, Universitat de
        Barcelona, Av.~Diagonal 647, \mbox{08028 Barcelona}, Spain}
\date{\today}
\maketitle

\begin{abstract}
In the first part of this paper, we
show that the semiclassical Einstein-Langevin equation, 
introduced in the framework of a stochastic
generalization of semiclassical gravity 
to describe the back reaction of matter 
stress-energy fluctuations, can be formally derived from a functional
method based on the influence functional of Feynman and Vernon. 
In the second part, we derive a number of results for 
background solutions of semiclassical gravity consisting of
stationary and conformally stationary spacetimes
and scalar fields in thermal equilibrium states. 
For these cases, fluctuation-dissipation relations are derived.
We also show that particle creation
is related to the vacuum stress-energy fluctuations
and that it is enhanced by the presence of 
stochastic metric fluctuations. 
\end{abstract}


%\narrowtext

%%%%%%%%%%%%%%%%%%%%%%%%%%%%%%%%%%%%%%%%

\section{Introduction}
\label{sec:introduction}

%%%%%%%%%%%%%%%%%%%%%%%%%%%%%%%%%%%%%%%%

It is generally believed that there must be a regime 
in which the gravitational field can be treated as a
classical or ``quasiclassical'' field, but its interaction with
quantum matter fields cannot be neglected. 
The standard approach to describe such a regime is the
semiclassical theory of gravity based on the semiclassical
Einstein equation. This is a generalization of the Einstein
equation for a classical metric when the expectation value of the
stress-energy tensor of quantum matter fields is the source of
curvature. The semiclassical theory of gravity is mathematically
consistent and fairly well understood, at least for linear
matter fields \cite{wald94,wald77,fulling,birrell,mostepanenko}.

One expects that semiclassical gravity could be derived 
as an approximation of 
a fundamental quantum theory of gravity. 
However, in the absence of such a fundamental theory, the scope and
limits of the semiclassical theory are less well understood
\cite{wald94,flanagan}. 
It has been pointed out, nevertheless, that this semiclassical theory
may not be valid when the matter fields have important quantum
stress-energy 
fluctuations \cite{wald94,wald77,birrell,ford82}.
When this is the case, the stress-energy
fluctuations may have relevant back-reaction effects on the spacetime
geometry in the form of induced gravitational
fluctuations \cite{ford82}.
A number of examples have been studied, both in cosmological
and in flat spacetimes, where, for some states 
of the matter fields, the stress-energy tensor have significant
fluctuations \cite{kuo93}. 
It is thus necessary to extend the semiclassical theory of
gravity to determine the effect of such fluctuations.

To address this problem, different approaches have been
adopted. The aim of the first part of the present paper is to
unify two of these approaches.

One of these approaches relies on the idea, first proposed
by Hu \cite{hu89} in the context of semiclassical cosmology,
of viewing the metric field as the ``system'' of interest and the
matter fields as being part of its ``environment.''
This approach leads naturally to 
the influence functional formalism of Feynman and
Vernon \cite{feynman-vernon}.  
In this formalism, the integration of the environment
variables in a path integral yields the influence functional, from
which one can define an effective action for the dynamics of the
system \cite{feynman-hibbs,calzettahu,humatacz,husinha,%
caldeira,hu-paz-zhang,hu-matacz94,greiner}.
This approach has been extensively used in the
literature, not only in the framework of semiclassical cosmology
\cite{calzettahu,humatacz,husinha,cv96,lomb-mazz,cv97,ccv97,%
campos-hu,campos-hu2,calver98}, but
also in the context of analogous semiclassical regimes in
quantum mechanics \cite{caldeira,hu-matacz94,hu-paz-zhang2}
and in quantum field theory 
\cite{greiner,matacz,morikawa,shaisultanov,gleiser}.
It is based on the observation that the semiclassical equation can be
directly  
derived from the effective action of Feynman and Vernon
\cite{calzettahu,greiner,cv96,ccv97,campos-hu,shaisultanov}.
When computing this effective action perturbatively up to quadratic
order in its variables, one usually finds some imaginary terms which
do not contribute to the semiclassical equation. 
The key point is then to formally identify the contribution of such
terms in the influence functional with the characteristic functional of
a Gaussian stochastic source. Assuming that, in the semiclassical
regime, this stochastic source interacts with the system variables,
equations of the
Langevin type can be derived for these variables. 
However, since this approach relies on a purely formal identification,
doubts may be raised on
the physical significance of the derived equations.

An alternative approach has been introduced in a recent
paper \cite{mv98}. In that work, we proposed a stochastic
semiclassical theory of gravity as a perturbative generalization of
semiclassical gravity to describe the back reaction of the lowest
order stress-energy fluctuations. The idea is in fact quite simple.
One starts realizing that, for 
a given solution of semiclassical gravity, the lowest order matter
stress-energy fluctuations can be associated to a
classical stochastic tensor field. 
Then, we seek an equation which incorporates in a consistent way 
this stochastic tensor as the source of linear perturbations to the
semiclassical metric.
The resulting equation is the semiclassical Einstein-Langevin
equation.

We should emphasize that, even if the metric fluctuations in this 
theory are classical (stochastic fluctuations), their
origin is presumably quantum. This is so not only because these metric
fluctuations are induced by the fluctuations of a quantum
operator, but also because they are supposed to describe some remnants
of the quantum gravity fluctuations after some mechanism for
decoherence and classicalization of the metric field
\cite{gell-mann-hartle,hartle,dowker,halliwell,whelan}.
From the formal assumption that such a mechanism is the Gell-Mann
and Hartle mechanism of environment-induced decoherence 
of suitably coarse-grained system variables 
\cite{gell-mann-hartle,hartle},
one may, in fact, derive the stochastic semiclassical theory 
\cite{mv98_2}.
Nevertheless, that derivation is of course formal,
given that, due to the lack of the full quantum theory of gravity, the
classicalization mechanism for the gravitational field
is not understood. 
One expects that the stochastic semiclassical theory is valid
when the characteristic time and space scales of variation 
of the metric field are well above
its characteristic decoherence scales.
In this regime, the theory can be applied 
to compute correlation functions of gravitational perturbations for
points separated by scales larger than these 
decoherence scales.  
Hence, this theory 
may have a number of interesting applications in black hole physics
and in cosmology,
particularly in view of the problem of structure
formation. Some examples of simple applications have already been
given in Refs.~\cite{ccv97,calver98,mv98}.

The purpose of the second part of the paper is to derive 
some general results concerning stochastic semiclassical 
gravity for stationary and conformally stationary background solutions
of semiclassical gravity (for conformal matter fields in the latter
case). We analyze two issues: the existence of a
fluctuation-dissipation relation and the creation of particles by
stochastic metric perturbations.

Under very general conditions,
a fluctuation-dissipation relation is known to exist in 
models of quantum mechanics, and also in some models of quantum
many-body systems or quantum fields 
in the presence of classical fields
\cite{landau,kubo,grabert,schwinger61,weber,kubo85,martin,weber,%
jackiw}.
This is a relation between quantum fluctuations of a system
in a state of thermal equilibrium and the dissipative properties of
this system caused by classical linear perturbations on it. 
The idea of a fluctuation-dissipation relation in the theory of
quantum fields in curved spacetimes and in the semiclassical
back-reaction problem was already present in some
early papers \cite{sciama,mottola,hu89}. 
A fluctuation-dissipation relation has been found 
in some of the previous derivations of semiclassical 
Langevin-type equations
\cite{husinha,cv96,campos-hu,campos-hu2}. 
Some authors believe that such a relation should always be
present and embody the physics of the back reaction of matter fields
on the gravitational field 
\cite{husinha,campos-hu,campos-hu2,hu99,hu-raval-sinha}.
It is also believed that noise and dissipation must be
related to the creation of particles by stochastic metric
perturbations 
\cite{hu89,calzettahu,humatacz,husinha,cv97,ccv97,hu99,hu-raval-sinha}.

In stationary and conformally stationary spacetimes  
(for conformal fields in the latter case), one can define a state of
thermal equilibrium for the matter fields. 
When the background solution of semiclassical
gravity is of one of these types, we can identify a dissipation
kernel in the corresponding semiclassical Einstein-Langevin equation
which is related to the fluctuations of the stochastic source by a
fluctuation-dissipation relation. 
We also study the production of
particles by stochastic metric perturbations to such backgrounds:
we relate particle
creation to the vacuum stress-energy fluctuations and 
we show that the
mean value of created particles is enhanced by the 
presence of metric fluctuations.

The plan of the paper is the following. 
In Sec.~\ref{sec:E-L}, we construct the stochastic semiclassical
theory of gravity to describe the back reaction of the stress-energy
fluctuations on the spacetime.    
In Sec.~\ref{sec:influence action}, we show that the 
semiclassical Einstein-Langevin equation obtained in 
Sec.~\ref{sec:E-L} can actually be formally derived with the 
functional approach. This connection 
clarifies the physical meaning of the 
Langevin-type equations previously derived by  
functional methods
\cite{calzettahu,humatacz,husinha,cv96,lomb-mazz,cv97,ccv97,%
campos-hu,campos-hu2,calver98},
since it shows that the formally introduced stochastic source
is directly related to the matter stress-energy fluctuations.
We then use the functional approach to write the Einstein-Langevin
equation in an explicit form, which is more suitable 
for specific calculations. 
In Sec.~\ref{sec:stationary}, we derive the 
fluctuation-dissipation relation for stationary and conformally
stationary backgrounds and the results for particle creation by
stochastic metric perturbations. 
Finally, in Sec.~\ref{sec:conclu}, we summarize our main conclusions.

Throughout this paper we use the $(+++)$ sign conventions and the
abstract index notation of Ref.~\cite{wald84},
and we work with units in which $c=\hbar =1$.

%%%%%%%%%%%%%%%%%%%%%%%%%%%%%%%%%%%%%%%%

\section{Stochastic semiclassical 
gravity}
\label{sec:E-L}

%%%%%%%%%%%%%%%%%%%%%%%%%%%%%%%%%%%%%%%%

In this section, we construct the stochastic semiclassical
theory of gravity as a perturbative extension of semiclassical
gravity to describe the back reaction of quantum stress-energy
fluctuations on the gravitational field.
Let us begin with a brief overview of the semiclassical
theory of gravity interacting with linear matter fields. 
Let $({\cal M},g_{ab})$ be a globally hyperbolic four-dimensional
spacetime and consider a linear 
quantum field $\Phi$ on it. For the sake of definiteness, we will
take $\Phi$ as a real scalar field, but all the analysis of this
section is valid for any kind of linear quantum field or
for a set of linear independent quantum fields. Throughout this
section we shall work in the Heisenberg picture. 
The field operator in this picture, $\hat{\Phi}$, is an
operator-valued distribution solution of 
the Klein-Gordon equation,   
\be
\left( \Box -m^2- \xi R \right) \hat{\Phi}=0,
\label{Klein-Gordon}
\ee
where $m$ is the mass,
$\Box \!\equiv\! \bigtriangledown_{\!a}\bigtriangledown^{a}$,
with $\bigtriangledown_{\!a}$ being the covariant derivative
associated to 
the metric $g_{ab}$, and $\xi$ is a dimensionless parameter coupling 
the field to the scalar curvature $R$. To indicate that the field
operator is a functional of the metric $g_{ab}$, we will 
write $\hat{\Phi}[g](x)$.

The classical stress-energy tensor is
obtained by functional derivation of
the classical action for the field in a background
spacetime $({\cal M},g_{ab})$ 
with respect to the metric. This tensor is a functional 
$T_{ab}[g,\Phi]$ of the metric $g_{ab}$ and of the classical field
$\Phi$. For a real scalar field, it is
\be
T_{ab}[g,\Phi]=\bigtriangledown_{\!a}\Phi
\bigtriangledown_{\!b}\!\Phi- {1\over 2}\, g_{ab} 
\bigtriangledown^{c}\!\Phi \bigtriangledown_{\!c}\!\Phi 
-{1\over 2}\, g_{ab}\, m^2 \Phi^2 
+\xi \left( g_{ab} \Box
-\bigtriangledown_{\!a}\!\! \bigtriangledown_{\!b}
+\, G_{ab} \right) \Phi^2,
\label{class s-t} 
\ee
where $G_{ab}$ is the Einstein tensor.
The next step is to define a stress-energy tensor operator 
$\hat{T}_{ab}[g](x)$. In a naive way, one would replace 
the classical field $\Phi$ in the functional
$T_{ab}[g,\Phi]$ by its corresponding quantum operator 
$\hat{\Phi}[g]$. However, since the field
operator is well-defined only as a distribution on spacetime and this 
procedure involves taking the product of two distributions at the same
spacetime point, the formal expression for
$\hat{T}_{ab}[g]$ is ill-defined and we need a regularization
procedure. 
We may formally think of a regularized ``operator'' 
$\hat{T}_{ab}[g](x;\Omega)$, 
depending on some regulator $\Omega$, defined by giving 
a precise prescription for computing its 
matrix elements for physically acceptable states of the field.
These states are assumed to be Hadamard states on the
Fock space of a Hadamard vacuum state \cite{wald94}.
The states may have to be regularized also in some way and the
procedure may involve some analytic continuation in the values of the
regulator. 
Of course, if we remove the regularization in the results for these
matrix elements, we would obtain infinite quantities.

Once the regularization prescription has been introduced,  
a renormalized and regularized stress-energy
``operator'' $\hat{T}^{R}_{ab}[g](x;\Omega)$ may be defined as 
\be
\hat{T}^{R}_{ab}[g](x;\Omega)=\hat{T}_{ab}[g](x;\Omega)+
F^{C}_{ab}[g](x;\Omega)\, \hat{I},
\label{renorm s-t} 
\ee
where $\hat{I}$ is the identity operator and $F^{C}_{ab}[g]$ 
are some symmetric tensor counterterms, 
which can be written in terms of the regulator $\Omega$ and local
functionals of the metric $g_{cd}(x)$.\footnote{In the 
point-splitting regularization method, for instance, 
one introduces a point $y$ in a normal neighborhood of the point $x$,
so some non-local dependence on the metric is explicitly introduced in
the regularized stress-energy operator and then also in the
counterterms. Using this regularization technique, the  
regulator can be taken as the vector $\sigma^a(x,y)$, which is the
tangent vector at the point $x$ to the geodesic joining $x$ and $y$
with length equal to the arc length along this geodesic. In this case,
the counterterms can be written in terms of the vector $\sigma^a(x,y)$
and tensors which are local  
functionals of the metric $g_{ab}(x)$ \cite{fulling,christensen}.}
These counterterms can and must be chosen in such a way
that, for any pair of physically acceptable states
$|\psi\rangle$ and $|\varphi\rangle$, the matrix element
of the renormalized operator $\hat{T}^{R}_{ab}[g]$, defined by
\be
\langle\psi|\hat{T}^{R}_{ab} |\varphi\rangle \equiv 
\lim_{\Omega \rightarrow \Omega_p} 
\langle\psi|\hat{T}^{R}_{ab} |\varphi\rangle (\Omega),
\label{renorm s-t 2}
\ee
where $\Omega_p$ means the ``physical value'' of the regulator, 
is finite (well defined as a distribution) and satisfies Wald's
axioms \cite{fulling,wald77}.
Using the point-splitting
or the dimensional regularization methods, these
counterterms can be extracted from the singular part of a
Schwinger-DeWitt series \cite{fulling,christensen,bunch}. 
The choice
of these counterterms is not unique, each different choice is
called a ``renormalization scheme,'' and this leads to some ambiguity
in the definition of the renormalized stress-energy tensor operator. 
But this ambiguity can be 
absorbed into the renormalized coupling constants
appearing in the equations of motion for the gravitational field. 
Thus, the ambiguity is only a mathematical artifact of the separation
of the action into a gravitational part and a matter part, but the
physically relevant equations are in fact unique 
\cite{fulling,fulling74}.

The semiclassical Einstein equation for the metric $g_{ab}$ can then 
be written as 
\be
{1\over 8 \pi G} \left( G_{ab}[g]+ \Lambda g_{ab} \right)-
2  \left( \alpha A_{ab}+\beta B_{ab} \right)\hspace{-0.3ex}[g]=
\langle \hat{T}^{R}_{ab} \rangle [g], 
\label{semiclassical Einstein eq}
\ee
where $\langle\hat{T}^{R}_{ab} \rangle [g]$ is the
expectation value of $\hat{T}^{R}_{ab}[g]$
in some physically acceptable state of the quantum field on the
spacetime $({\cal M},g_{ab})$. The notation 
$\langle\hat{T}^{R}_{ab}\rangle [g]$ is used 
to indicate that this expectation value is a functional of the metric
$g_{cd}$, not only because the stress-energy tensor operator depends
on the metric, but also because the state of the matter field 
depends on the spacetime
(in general, such
state depends on the global structure of the spacetime manifold).
In Eq.~(\ref{semiclassical Einstein eq}), 
$G$, $\Lambda$, $\alpha$ and
$\beta$ are renormalized coupling constants, respectively, the
Newtonian gravitational constant, the cosmological constant and two
dimensionless coupling constants. These constants may be seen 
as the result of ``dressing'' the bare coupling constants in a
suitably regularized version of the gravitational part of the action, 
\be
S_{g}[g] \equiv 
\int \! d^4 x \, \sqrt{- g} \left[{1\over 16 \pi G_{B}}
\left(R-2\Lambda_{B}\right)
+ \alpha_{B} C_{abcd}C^{abcd}+\beta_{B} R^2 \right],
\label{grav action}
\ee
where $C_{abcd}$ is the Weyl tensor and the subindex 
${\scriptstyle B}$ in the
coupling constants means ``bare.'' 
These renormalized coupling constants are supposed to be determined
experimentally (for the specific renormalization scheme that one has
chosen and for the characteristic scales of the physics under
consideration). 
The tensors $A_{ab}$ and $B_{ab}$ in 
Eq.~(\ref{semiclassical Einstein eq}) come from the functional
derivatives with 
respect to the metric of the terms
quadratic in the curvature in $S_{g}[g]$, which are needed to ensure
the renormalizability of the theory. These tensors are explicitly
given by
\begin{eqnarray}
A^{ab} \equiv {1\over\sqrt{- g}}   \frac{\delta}{\delta g_{ab}}
                \int\! d^4 x \, \sqrt{- g}\, C_{cdef}C^{cdef}
            &=& {1\over2}\,g^{ab}C_{cdef} C^{cdef}
                -2R^{acde}{R^b}_{cde}
                +4R^{ac}{R_c}^b  \nonumber \\
            && -\,{2\over3}\,RR^{ab}
               -2\hspace{0.2ex} \Box \hspace{-0.2ex} R^{ab}
               +{2\over3} \bigtriangledown^{a}\!\bigtriangledown^{b} R
               +{1\over3}\,g^{ab} \hspace{0.2ex} \Box \hspace{-0.2ex}  R,
\label{A}
\end{eqnarray}
and
\be
\!\!\!\!\!\!\!\!\!\!\!\!\!\!\!\!\!\!\!\!\!\!\!\!\!\!\!\!
B^{ab} \equiv {1\over\sqrt{- g}}   \frac{\delta}{\delta g_{ab}}
 \int\! d^4 x\,\sqrt{- g}\, R^2 
 = {1\over2}\,g^{ab} R^2-2 R R^{ab}
+2 \bigtriangledown^{a}\!\bigtriangledown^{b} R
   -2 g^{ab} \hspace{0.2ex} \Box \hspace{-0.2ex} R,
\label{B}           
\ee
where $R_{abcd}$ is the Riemann tensor and $R_{ab}$ is the Ricci tensor.
Note that each of the terms in Eq.~(\ref{semiclassical Einstein eq})
has vanishing divergence.
Notice also that we could add a classical stress-energy
tensor to the right hand side of 
Eq.~(\ref{semiclassical Einstein eq}), if we had a classical matter
source, but, for simplicity, we shall ignore such a term.

As long as the gravitational field is assumed to be described by a
classical Lorentzian metric $g_{ab}$, the semiclassical
Einstein equation seems to be the only physically
plausible dynamical equation for this metric.  
The reason is that, in classical general relativity, the metric
$g_{ab}$ couples to matter through the stress-energy 
tensor. For a field quantized on the spacetime $({\cal M},g_{ab})$ 
and for a given state of this field, the expectation value of the 
renormalized stress-energy tensor operator is the only physically
observable (up to the ambiguity mentioned above) c-number
stress-energy tensor that we can construct.

A solution of semiclassical gravity consists of a
spacetime $({\cal M},g_{ab})$, a quantum field operator 
$\hat{\Phi}[g]$ satisfying Eq.~(\ref{Klein-Gordon}), 
and a physically acceptable state $|\psi\rangle[g]$ 
for this field (which can also be a mixed state characterized by a 
density operator), 
such that Eq.~(\ref{semiclassical Einstein eq}) is
satisfied when the expectation value in the state $|\psi\rangle[g]$
of the renormalized operator $\hat{T}^{R}_{ab}[g]$
is put on the right hand side.

Let us now introduce stress-energy fluctuations.
Given a solution of semiclassical gravity, the stress-energy
tensor will in general have quantum fluctuations. To lowest order,
such fluctuations are described by the bi-tensor, which shall be
called noise kernel, defined by
\be
8 N_{abcd}(x,y) \equiv \lim_{\Omega \rightarrow \Omega_p}
\bigl\langle  \bigl\{
 \hat{t}_{ab}(x) , \,
 \hat{t}_{cd}(y)
 \bigr\} \bigr\rangle [g](\Omega),
\label{noise}
\ee
where $\{ \; , \: \}$ means the anticommutator and
$\hat{t}_{ab}(x;\Omega) \equiv \hat{T}_{ab}(x;\Omega)-
 \langle \hat{T}_{ab}(x) \rangle (\Omega)$.
Note that we have
defined this noise kernel in terms of the unrenormalized  
``operator'' $\hat{T}_{ab}[g](x;\Omega)$. 
For a linear quantum field, this can be done
because the ultraviolet singular
behavior of $\langle\hat{T}_{ab}(x)
\hat{T}_{cd}(y)\rangle (\Omega)$ is the same as that of 
$\langle\hat{T}_{ab}(x)\rangle (\Omega)
\langle\hat{T}_{cd}(y) \rangle (\Omega)$, so 
$N_{abcd}(x,y)$ is free of ultraviolet divergencies. 
One can trivially see from the substitution of (\ref{renorm s-t})
into (\ref{noise}) that
we can replace $\hat{T}_{ab}[g](x;\Omega)$ by
the renormalized operator $\hat{T}^{R}_{ab}[g](x)$,
and omit the limit 
$\Omega \!\rightarrow \! \Omega_p$, in the last expression.
The result is obviously
independent of the renormalization scheme that one chooses
to define $\hat{T}^{R}_{ab}$.

As a perturbative correction to semiclassical gravity, 
we want now to introduce an equation in which the 
stress-energy fluctuations described by (\ref{noise}) 
are the source of classical gravitational fluctuations.
Thus, we assume that the gravitational field is 
described by  $g_{ab}+h_{ab}$,
where $h_{ab}$ is a linear perturbation 
to the background metric $g_{ab}$, solution of
Eq.~(\ref{semiclassical Einstein eq}). 
The renormalized stress-energy
operator and the state of the quantum field
may be denoted by $\hat{T}^{R}_{ab}[g+h]$ and
$|\psi\rangle[g+h]$, respectively, and 
$\langle\hat{T}^{R}_{ab} \rangle [g+h]$
is the corresponding expectation value.

Let us introduce a Gaussian stochastic tensor field $\xi_{ab}$
defined by the following correlators:
\be
\langle\xi_{ab}(x) \rangle_c = 0,  
\hspace{6ex}
\langle\xi_{ab}(x)\xi_{cd}(y) \rangle_c = N_{abcd}(x,y),
\label{correlators}
\ee
where 
$\langle \hspace{1.5ex} \rangle_c$ means statistical 
average.
In general, the two-point correlation function of a stochastic tensor
field $\xi_{ab}$ must be a symmetric,
in the sense that 
$\langle\xi_{ab}(x)\xi_{cd}(y) \rangle_c =
\langle\xi_{cd}(y)\xi_{ab}(x) \rangle_c$, and
positive semi-definite real bi-tensor field. 
Since the renormalized operator $\hat{T}^{R}_{ab}$ is  
self-adjoint,
it is easy to see from the definition
(\ref{noise}) that $N_{abcd}(x,y)$ satisfies all these conditions.
Therefore, the relations (\ref{correlators}), with the cumulants of
higher order taken to be zero, do truly characterize a stochastic
tensor field $\xi_{ab}$.
The simplest equation which can incorporate in a consistent way the
stress-energy fluctuations described by $N_{abcd}(x,y)$
as the source of classical metric fluctuations is
\be
{1\over 8 \pi G} \Bigl( G_{ab}[g+h]+ 
\Lambda\left(g_{ab}+h_{ab}\right) \Bigr)- 
2 \left( \alpha A_{ab}+\beta B_{ab} \right)\hspace{-0.3ex}
[g+h]=\langle
\hat{T}^{R}_{ab}\rangle [g+h] +2 \xi_{ab} , 
\label{Einstein-Langevin eq}
\ee 
%\bea
%&&{1\over 8 \pi G} \Bigl( G_{ab}[g+h]+ 
%\Lambda\left(g_{ab}+h_{ab}\right) \Bigr)    \nn \\
%&& \hspace{3.5ex}
%-\,2 \left( \alpha A_{ab}+\beta B_{ab} \right)\hspace{-0.3ex}
%[g+h]=\langle
%\hat{T}^{R}_{ab} \rangle [g+h] +2 \xi_{ab}, 
%\label{Einstein-Langevin eq}
%\eea
which must be understood as a dynamical equation for $h_{ab}$ 
to linear order. 
Eq.~(\ref{Einstein-Langevin eq}) 
is the semiclassical Einstein-Langevin equation, which 
gives a first order correction to semiclassical gravity.
One could also seek equations describing higher order corrections, 
which would involve higher order stress-energy fluctuations, but,
for simplicity, we shall stick to the lowest order.

In order to check the consistency of 
Eq.~(\ref{Einstein-Langevin eq}), note that 
the term $\xi_{ab}$ does not depend on $h_{cd}$,
since it is completely determined from the solution of
semiclassical gravity by the correlators
(\ref{correlators}). Even so, this term must be considered as 
of first order in perturbation theory around
semiclassical gravity. As shown in Ref.~\cite{mv98},
$\xi_{ab}$
is covariantly conserved up to first order in this perturbation
theory, in the sense that $\bigtriangledown^{a} \xi_{ab}$
behaves deterministically as the zero vector field 
on ${\cal M}$ 
($\bigtriangledown^{a}$ is the covariant derivative 
associated to the background metric $g_{ab}$).
It is thus consistent to include the
term $\xi_{ab}$ in the right hand side of 
Eq.~(\ref{Einstein-Langevin eq}).

It was also shown in Ref.~\cite{mv98} that
for a conformal field, {\it i.e.}, a field whose classical action is
conformally invariant ({\it e.g.}, a massless conformally coupled
scalar field), $\xi_{ab}$
is ``traceless'' up to first order in perturbation theory,
since $g^{ab}\xi_{ab}$ behaves deterministically as a vanishing
scalar. 
Hence, in the case of a conformal matter field, 
the trace of 
the right hand side of Eq.~(\ref{Einstein-Langevin eq}) 
comes only from the trace anomaly.

Since Eq.~(\ref{Einstein-Langevin eq}) is a linear stochastic
equation for $h_{ab}$ with an inhomogeneous term $\xi_{ab}$,
a solution can be formally written as a functional
$h_{ab}[\xi]$. 
Such a solution can be characterized by 
the whole family of its correlation functions. 
From the average of
Eq.~(\ref{Einstein-Langevin eq}), the average of the
metric, $g_{ab}+\langle h_{ab} \rangle_c$,
must be a solution of the semiclassical Einstein equation linearized
around $g_{ab}$. The fluctuations of the metric around this average
can be described by the moments of order higher than one of the
stochastic field
$h_{ab}^{\rm f}[\xi] \equiv h_{ab}[\xi]
-\langle h_{ab} \rangle_c$.

Finally, for the solutions of Eq.~(\ref{Einstein-Langevin eq}) 
we have the gauge freedom 
$h_{ab} \rightarrow 
h'_{ab}\equiv h_{ab}+ \bigtriangledown_{\!a}
\zeta_{b}+\bigtriangledown_{\!b} \zeta_{a}$, 
where $\zeta^{a}$ is any stochastic vector field on ${\cal M}$ which
is a functional of $\xi_{cd}$, 
and $\zeta_{a} \equiv g_{ab}\zeta^{b}$.
Note that the tensors which appear in 
Eq.~(\ref{Einstein-Langevin eq}) transform as 
$R_{ab}[g\!+\!h']\!=\! R_{ab}[g\!+\!h]\!+\!
{\cal \pounds}_{\mbox{}_{\! \zeta}} R_{ab}[g]$ 
(to linear order in the perturbations),
where ${\cal \pounds}_{\mbox{}_{\! \zeta}}$ is the Lie 
derivative with respect to $\zeta^a$.
If we substitute $h_{ab}$ by $h'_{ab}$ in
Eq.~(\ref{Einstein-Langevin eq}), we get 
Eq.~(\ref{Einstein-Langevin eq}) plus the Lie derivative 
of a combination of the tensors which appear in 
Eq.~(\ref{semiclassical Einstein eq}). This last tensorial
combination vanishes when Eq.~(\ref{semiclassical Einstein eq}) is
satisfied. Thus, it is necessary that the 
set $({\cal M},g_{ab},\hat{\Phi}[g],|\psi\rangle[g])$ be a solution of
semiclassical gravity to ensure that the Einstein-Langevin equation
(\ref{Einstein-Langevin eq}) is gauge invariant.

%%%%%%%%%%%%%%%%%%%%%%%%%%%%%%%%%%%%%%%%

\section{Derivation from an influence action}
\label{sec:influence action}

%%%%%%%%%%%%%%%%%%%%%%%%%%%%%%%%%%%%%%%%

The purpose of this section is to derive the semiclassical 
Einstein-Langevin equation (\ref{Einstein-Langevin eq}) by
a method based on functional techniques. 
The same method has been in fact used  
in the literature to derive
Langevin-type equations in the context of semiclassical cosmology 
\cite{calzettahu,humatacz,husinha,cv96,lomb-mazz,cv97,ccv97,campos-hu,%
campos-hu2,calver98}
and of analogous semiclassical regimes for systems of
quantum mechanics \cite{caldeira,hu-matacz94,hu-paz-zhang2}
and of quantum field theory 
\cite{greiner,matacz,morikawa,shaisultanov,gleiser}.
Using these functional techniques, we also work out the
Einstein-Langevin equation 
more explicitly, in a form more suitable for specific
calculations. Here, we consider again 
the simplest case of a linear real scalar field $\Phi$.

These functional techniques are based on 
the closed time path (CTP) functional formalism, due 
to Schwinger and Keldysh \cite{schwinger61,schwinger62}.
This formalism is designed to obtain expectation
values of field operators in a direct way
and it is suited to derive dynamical equations for expectation values;
see Refs.~\cite{ctp,cv94,campos-hu} for detailed reviews. 
In our case, this formalism will be useful to obtain an
expression for the expectation value 
$\langle \hat{T}^{ab}\rangle [g\!+\!h]$ as an expansion in
the metric perturbation.  
When the full quantum system consists of a distinguished subsystem 
(the ``system'' of interest)
interacting with an environment (the remaining degrees of freedom),
the CTP functional formalism  
turns out to be related 
\cite{calzettahu,greiner,cv96,campos-hu,morikawa,shaisultanov,mv98_2}
to the
influence functional formalism of Feynman and Vernon
\cite{feynman-vernon}. 
In this latter formalism, the integration of the environment
variables in a CTP path integral yields the influence functional, from
which one can define an effective action for the dynamics of the
system \cite{feynman-hibbs,calzettahu,humatacz,husinha,%
caldeira,hu-paz-zhang,hu-matacz94,greiner}. 
Applying this influence functional formalism to our problem, the
semiclassical Einstein-Langevin equation will be formally derived in
subsection \ref{subsec:formal derivation}.

In our case, we consider
the metric field $g_{ab}(x)$ as the ``system'' degrees of freedom, 
and the scalar field $\Phi(x)$ 
and also some ``high-momentum'' gravitational modes
\cite{whelan} as the ``environment'' variables. 
Unfortunately, since the form of a complete
quantum theory of gravity interacting with matter is unknown,
we do not know what these ``high-momentum'' gravitational modes are.
Such a fundamental quantum theory might not even be a field theory,
in which case the metric and scalar fields would not be 
fundamental objects \cite{hu99}.
Thus, in this case, we cannot attempt to evaluate 
the influence action of Feynman and Vernon 
starting from the fundamental quantum theory and 
performing the path integrations in the environment variables.
Instead, we introduce the influence action for 
an effective quantum field theory of gravity and matter
\cite{donoghue,humatacz}, 
in which such ``high-momentum'' gravitational
modes are assumed to have been already ``integrated out.'' 
Adopting the usual procedure of 
effective field theories \cite{weinberg,donoghue},
one has to take the effective action for the metric and the scalar
field of 
the most general local form compatible with general covariance: 
$S[g,\Phi] \!\equiv \! S_g[g]+S_m[g,\Phi]+ \cdots$, where $S_g[g]$ 
is given by (\ref{grav action}), 
\be
S_m[g,\Phi] \equiv -{1\over2} \int\! d^4x \, \sqrt{- g} 
  \left[g^{ab}\partial_a \Phi \hspace{0.2ex} \partial_b \Phi
  +\left(m^2+ \xi R \right)\Phi^2 \right], 
\label{scalar field action}
\ee
and the dots  
stand for terms of order higher than two 
in the curvature and in the number of
derivatives of the scalar field [because of the
Gauss-Bonnet theorem in four spacetime dimensions, 
no further terms of second order in the curvature are needed in the
gravitational action (\ref{grav action})].
In this paper, we shall neglect the higher order terms as well as 
self-interaction terms for the scalar field.
The second order terms are necessary to renormalize 
one-loop ultraviolet divergencies of the scalar field 
stress tensor. 
Since ${\cal M}$ is a globally hyperbolic manifold, 
we can foliate it by a family of $t\!=\! {\rm constant}$ Cauchy
hypersurfaces $\Sigma_{t}$. We denote by  
${\bf x}$ the coordinates on each of these hypersurfaces, and by
$t_{i}$ and $t_{f}$ some initial and final times, respectively.
The integration domain for the action terms must be understood
as a compact region ${\cal U}$ of the manifold ${\cal M}$, bounded by
the hypersurfaces $\Sigma_{t_i}$ and $\Sigma_{t_f}$.

Assuming the form (\ref{scalar field action}) for 
the effective action which couples the scalar and the metric fields,  
we can now
introduce the corresponding influence functional.
This is a functional of two copies of the metric field that we denote
by $g_{ab}^+$ and $g_{ab}^-$.
Let us assume that, in the quantum effective theory, 
the state of the full system 
(the scalar and the metric fields) in the Schr\"{o}dinger picture 
at the initial time $t\! =\! t_{i}$ can be described by 
a factorizable 
density operator, {\it i.e.}, a density operator which can be written
as the tensor product of two operators on the Hilbert spaces
of the metric and of the scalar field.
Let $\hat{\rho}^{\rm \scriptscriptstyle S}(t_{i})$ be the 
density operator describing the initial state of the 
scalar field.
If we consider the theory of a scalar field quantized in the
classical background spacetime $({\cal M},g_{ab})$ through the action 
(\ref{scalar field action}), a
state in the Heisenberg picture described by a density operator
$\hat{\rho}[g]$ corresponds to this state.
Let $\left\{ |\varphi(\mbox{\bf x})\rangle^{\rm \scriptscriptstyle S} 
\right\}$ 
be the basis of eigenstates of the scalar
field operator $\hat{\Phi}^{\rm \scriptscriptstyle S}({\bf x})$
in the Schr\"{o}dinger picture:
$\hat{\Phi}^{\rm \scriptscriptstyle S}({\bf x})
\, |\varphi\rangle ^{\rm \scriptscriptstyle S}=
\varphi(\mbox{\bf x})
\, |\varphi\rangle^{\rm \scriptscriptstyle S}$.
The matrix elements of 
$\hat{\rho}^{\rm \scriptscriptstyle S}(t_{i})$ in this basis will be
written as 
$\rho_{i} \!\left[\varphi,\tilde{\varphi}\right] \equiv 
\mbox{}^{\rm \scriptscriptstyle S}
\langle \varphi|\,\hat{\rho}^{\rm \scriptscriptstyle S}(t_{i})
\, |\tilde{\varphi}\rangle^{\rm \scriptscriptstyle S}$. 
We can now introduce the influence functional as the following
path integral over two copies of the scalar field:
\be
{\cal F}_{\rm IF}[g^+,g^-] \equiv
\int\! {\cal D}[\Phi_+]\;
{\cal D}[\Phi_-] \;
\rho_i \!\left[\Phi_+(t_i),\Phi_-(t_i) \right] \,
\delta\!\left[\Phi_+(t_f)\!-\!\Phi_-(t_f)  \right]\:
e^{i\left(S_m[g^+,\Phi_+]-S_m[g^-,\Phi_-]\right) }.
\label{path integral}
\ee 
The above double path integral can be rewritten as a closed time path
(CTP) integral, namely, as an integral over a single copy of field
paths with two different time branches, one going forward in time from
$t_i$ to $t_f$, and the other going backward in time from $t_f$ 
to $t_i$. 
From this influence functional, 
the influence action, $S_{\rm IF}[g^+,g^-]$, and the effective action of
Feynman and Vernon, $S_{\rm eff}[g^+,g^-]$, are defined by
${\cal F}_{\rm IF}[g^+,g^-] \equiv
e^{i S_{\rm IF}[g^+,g^-]}$ and 
$S_{\rm eff}[g^+,g^-]\equiv S_{g}[g^+]-S_{g}[g^-]
+S_{\rm IF}[g^+,g^-]$.

Expression (\ref{path integral}) is ill-defined,
it must be regularized to get a meaningful 
influence functional. We shall assume that we can use 
dimensional regularization, that is, that we can give sense
to Eq.~(\ref{path integral}) by
dimensional continuation of all the quantities that appear in this
expression. 
We should point out, nevertheless, that 
for this regularization method to work one must be able to perform an
analytic continuation to Riemmanian signature
\cite{wald79}. 
Thus, we 
substitute the action $S_m$ in (\ref{path integral}) 
by some generalization to $n$ spacetime
dimensions, which may be chosen as 
\be
S_m[g,\Phi_{n}] = -{1\over2} \int\! d^n x \, \sqrt{- g} 
  \left[g^{ab} \partial_a \Phi_{n} \partial_b \Phi_{n}
  +\left(m^2+ \xi R \right)\Phi_{n}^2  \right],
\label{scalar action}
\ee  
where we use a notation in which a subindex $n$ is attached to 
these quantities that have different physical dimensions than the 
corresponding physical quantities in 
four dimensions. A quantity with the subindex $n$ can always
be associated to another without this subindex by means of a 
mass scale $\mu$; thus, for the scalar field
$\Phi_{n}\!=\! \mu^{(n-4)/2} \,\Phi$.

We also need to substitute the action (\ref{grav action}) by 
some suitable generalization to $n$ spacetime dimensions. We 
take
\be
S_g[g]=\mu^{n-4} \!\int \! d^n x \,\sqrt{- g} 
\left[{1\over 16 \pi G_{B}}
 \left(R-2\Lambda_{B}\right)+ {2\over 3}\,\alpha_{B} 
 \left(R_{abcd}R^{abcd}-
  R_{ab}R^{ab}  \right)+\beta_{B} R^2 \right].
\label{grav action in n}
\ee
By the Gauss-Bonnet theorem, this action gives for $n \!=\! 4$ 
the same equations of motion as the action (\ref{grav action}). 
The form of (\ref{grav action in n}) is suggested by the
Schwinger-DeWitt analysis of the ultraviolet divergencies in the
matter stress-energy tensor using dimensional regularization 
\cite{bunch}. 
Using (\ref{scalar action}) and (\ref{grav action in n}), one can
write the effective action of Feynman and Vernon, 
$S_{\rm eff}[g^+,g^-]$, in dimensional regularization. 
Since the action terms (\ref{scalar action}) and 
(\ref{grav action in n})
contain second order derivatives of the metric, one should also add
some boundary terms \cite{wald84,humatacz}. 
The effect of these
terms is to cancel out the boundary terms which appear
when taking variations of $S_{\rm eff}[g^+,g^-]$ keeping the value
of $g^+_{ab}$ and $g^-_{ab}$ fixed on the boundary of ${\cal U}$. 
Alternatively, in order to obtain the equations of motion for
the metric in the semiclassical regime, we can work with the action terms 
(\ref{scalar action}) and (\ref{grav action in n}) (without boundary
terms) and neglect all boundary terms when taking variations with
respect to $g^{\pm}_{ab}$. From now on, all the functional derivatives
with respect to the metric will be understood in this sense.

%%%%%%%%%%%%%%%%%%%%%%%%%%%%%%%%%%%%%%%%

\subsection{The semiclassical Einstein  
equation in dimensional regularization}
\label{subsec:semiclassical in n}

%%%%%%%%%%%%%%%%%%%%%%%%%%%%%%%%%%%%%%%%

From the action 
(\ref{scalar action}), we can define the stress-energy tensor
functional in the usual way 
\be
T^{ab}[g,\Phi_{n}](x) \equiv {2\over\sqrt{- g(x)}} \, 
   \frac{\delta S_m[g,\Phi_{n}]}{\delta g_{ab}(x)},
\label{s-t functional}
\ee
which yields (\ref{class s-t}). 
Working in the Heisenberg picture, we can now formally introduce the
regularized stress-energy tensor operator as 
\be
\hat{T}_{n}^{ab}[g] \equiv  T^{ab}[ g,\hat{\Phi}_{n}[g]], 
\hspace{5 ex}  
\hat{T}^{ab}[g] \equiv  \mu^{-(n-4)}\, \hat{T}_{n}^{ab}[g],
\label{regul s-t}
\ee
where $\hat{\Phi}_{n}[g](x)$ is the field operator, 
which satisfies the Klein-Gordon equation
(\ref{Klein-Gordon}) in $n$ spacetime dimensions,
and where we use a symmetrical ordering (Weyl ordering) prescription
for the operators. Using the Klein-Gordon equation, 
the stress-energy operator can be written as
\be
\hat{T}_{n}^{ab}[g] = {1\over 2} \left\{
     \bigtriangledown^{a}\hat{\Phi}_{n}[g]\, , \,
     \bigtriangledown^{b}\hat{\Phi}_{n}[g] \right\}
     + {\cal D}^{ab}[g]\, \hat{\Phi}_{n}^2[g],
\label{regul s-t 2}
\ee
where ${\cal D}^{ab}[g]$ is the differential operator
\be
{\cal D}^{ab}_{x} \equiv \left(\xi-{1\over 4}\right) g^{ab}(x) 
\Box_{x}+ \xi
\left( R^{ab}(x)- \bigtriangledown^{a}_{x} 
\bigtriangledown^{b}_{x} \right).
\label{diff operator}
\ee
From the definitions (\ref{path integral}),
(\ref{s-t functional}) and (\ref{regul s-t}), one
can see that 
\be
\langle \hat{T}_n^{ab}(x) \rangle [g] =
\left. {2\over\sqrt{- g(x)}} \, 
 \frac{\delta S_{\rm IF}[g^+,g^-]}
{\delta g^+_{ab}(x)} \right|_{g^+=g^-=g},
\label{s-t expect value}
\ee 
where the expectation value is taken in the $n$-dimensional spacetime
generalization of the state described by
$\hat{\rho}[g]$.
Therefore, differentiating 
$S_{\rm eff}[g^+,g^-]= S_{g}[g^+]-S_{g}[g^-]
+S_{\rm IF}[g^+,g^-]$ 
with respect to $g^+_{ab}$, and then setting
$g^+_{ab}=g^-_{ab}=g_{ab}$, we get
the semiclassical Einstein
equation in dimensional regularization:
\be
{1\over 8 \pi G_{B}} \left( G^{ab}[g]+ \Lambda_{B} g^{ab} \right)-
\left({4\over 3}\, \alpha_{B} D^{ab} 
+2  \beta_{B} B^{ab}\right)\! [g]
= \mu^{-(n-4)}
\langle \hat{T}_{n}^{ab}\rangle  [g], 
\label{semiclassical eq in n} 
\ee
where 
\bea
&&D^{ab} \equiv {1\over\sqrt{- g}}   \frac{\delta}{\delta g_{ab}}
          \int \! d^n x \,\sqrt{- g} \left(R_{cdef}R^{cdef}-
                                       R_{cd}R^{cd}  \right)
    = {1\over2}\, g^{ab} \! \left(  R_{cdef} R^{cdef}-
         R_{cd}R^{cd}+\Box \hspace{-0.2ex} R \right) 
      -2R^{acde}{R^b}_{cde}
\nn \\
&& \hspace{56.5ex}
      -\, 2 R^{acbd}R_{cd}
      +4R^{ac}{R_c}^b
      -3 \hspace{0.2ex}\Box \hspace{-0.2ex} R^{ab}
  +\bigtriangledown^{a}\!\bigtriangledown^{b}\! \hspace{-0.2ex} R,
\label{D}
\eea
and $B^{ab}$ is defined as in (\ref{B}) but 
for $n$ spacetime dimensions, although its explicit expression 
in terms of the metric and curvature tensors is the  
same. 
When $n\!=\!4$, one has that $D^{ab}\!=\!(3/2) A^{ab}$, 
where $A^{ab}$ is the tensor defined in (\ref{A}). 
From equation 
(\ref{semiclassical eq in n}), renormalizing the coupling
constants to eliminate the ``divergencies'' in 
$\mu^{-(n-4)} \langle \hat{T}_{n}^{ab}\rangle [g]$,
and taking the limit $n\!\rightarrow \! 4$, we
get the physical semiclassical Einstein equation 
(\ref{semiclassical Einstein eq}).

%%%%%%%%%%%%%%%%%%%%%%%%%%%%%%%%%%%%%%%%

\subsection{A formal derivation of 
the semiclassical Einstein-Langevin  
equation}
\label{subsec:formal derivation}

%%%%%%%%%%%%%%%%%%%%%%%%%%%%%%%%%%%%%%%%

In the spirit of the previous section, we now seek a dynamical
equation for a linear perturbation $h_{ab}$ to a semiclassical 
metric $g_{ab}$, solution of
Eq.~(\ref{semiclassical eq in n}) in $n$ spacetime dimensions.
From the result of the previous subsection, if such equation were
simply a linearized semiclassical Einstein equation, it could be
obtained from an expansion of the effective action
$S_{\rm eff}[g+h^+,g+h^-]$. In particular, since, from 
Eq.~(\ref{s-t expect value}), we have that
\be
\langle \hat{T}_n^{ab}(x) \rangle [g+h]=
\left. {2\over\sqrt{-\det (g\!+\!h)(x)}} \, 
 \frac{\delta S_{\rm IF}
   [g\!+\!h^+,g\!+\!h^-]}{\delta h^+_{ab}(x)} 
 \right|_{h^+=h^-=h},
\label{perturb s-t expect value}
\ee 
the expansion of $\langle \hat{T}_{n}^{ab}\rangle [g\!+\!h]$
to linear order in $h_{ab}$ can be obtained from an expansion of the
influence action $S_{\rm IF}[g+h^+,g+h^-]$ up to second order
in $h^{\pm}_{ab}$.

To perform the expansion of the influence action, 
we have to compute the first and second order
functional derivatives of $S_{\rm IF}[g^+,g^-]$
and then set $g^+_{ab}\!=\!g^-_{ab}\!=\!g_{ab}$.
If we do so using the path integral representation
(\ref{path integral}), we can interpret these derivatives as
expectation values of operators.
The relevant second order derivatives are
\bea
\left. {1\over\sqrt{- g(x)}\sqrt{- g(y)} } \,   
 \frac{\delta^2 S_{\rm IF}[g^+,g^-]}
{\delta g^+_{ab}(x)\delta g^+_{cd}(y)}
 \right|_{g^+=g^-=g} \!\!
&=& -H_{\scriptscriptstyle \!{\rm S}_{\scriptstyle n}}^{abcd}[g](x,y)
-K_n^{abcd}[g](x,y)+
i N_n^{abcd}[g](x,y),      \nn \\
\left. {1\over\sqrt{- g(x)}\sqrt{- g(y)} } \, 
 \frac{\delta^2 S_{\rm IF}[g^+,g^-]}
{\delta g^+_{ab}(x)\delta g^-_{cd}(y)} 
 \right|_{g^+=g^-=g} \!\!
&=& -H_{\scriptscriptstyle \!{\rm A}_{\scriptstyle n}}^{abcd}
[g](x,y)
-i N_n^{abcd}[g](x,y),  
\label{derivatives}  
\eea
where  
\bea
&&N_n^{abcd}[g](x,y) \equiv 
{1\over 8} \left\langle  \bigl\{
 \hat{t}_n^{ab}(x) , \,
 \hat{t}_n^{cd}(y)
 \bigr\} \right\rangle [g],
\hspace{8ex}
H_{\scriptscriptstyle \!{\rm S}_{\scriptstyle n}}^{abcd}
[g](x,y) \equiv 
{1\over 4}\:{\rm Im} \left\langle {\rm T}^{\displaystyle \ast}\!\!
\left( \hat{T}_n^{ab}(x) \hat{T}_n^{cd}(y) 
\right) \right\rangle \![g],       
\nn   \\
&&H_{\scriptscriptstyle \!{\rm A}_{\scriptstyle n}}^{abcd}
[g](x,y) \equiv 
-{i\over 8} \left\langle 
\bigl[ \hat{T}_n^{ab}(x), \, \hat{T}_n^{cd}(y)
\bigr] \right\rangle \![g],       
\hspace{5ex}
K_n^{abcd}[g](x,y) \equiv 
\left. {-1\over\sqrt{- g(x)}\sqrt{- g(y)} } \, \left\langle 
\frac{\delta^2 S_m[g,\Phi_{n}]}{\delta g_{ab}(x)\delta g_{cd}(y)} 
\right|_{\Phi_{n}=\hat{\Phi}_{n}}\right\rangle \![g],
\nn  \\
\mbox{}
\label{kernels}  
\eea  
with $\hat{t}_n^{ab} \equiv \hat{T}_{n}^{ab}-
\langle \hat{T}_{n}^{ab} \rangle$, and    
using again a Weyl ordering prescription
for the operators in the last of these expressions. 
Here, $[ \; , \: ]$ means the commutator, and we
use the symbol ${\rm T}^{\displaystyle \ast}$ 
to denote that, first,
we have to time order the field operators $\hat{\Phi}_{n}$ and then
apply the derivative operators which appear in each term 
of the product $T^{ab}(x) T^{cd}(y)$, where $T^{ab}$ is
the functional (\ref{class s-t}). For instance,
\be
{\rm T}^{\displaystyle \ast}\!\! \left(\hspace{-0.07ex} 
\bigtriangledown^{a}_{\!\!\! \mbox{}_{x}}
    \hspace{-0.1ex}\hat{\Phi}_{n}(x)\!
\bigtriangledown^{b}_{\!\!\! \mbox{}_{x}}\!\hat{\Phi}_{n}(x)\!
\bigtriangledown^{c}_{\!\!\! \mbox{}_{y}}\!\hat{\Phi}_{n}(y)\!
\bigtriangledown^{d}_{\!\!\! \mbox{}_{y}}\!\hat{\Phi}_{n}(y)\!
\right)\! =\!\!\!\!\lim_{ 
x_1,x_2 \rightarrow x_{\!\!\!\!\!\!\!\!\!\!\!\!\!\!\!\!\!\!\!\!\!
\!\!\!\!\!
\mbox{}_{\mbox{}_{\mbox{}_{\mbox{}_
{\mbox{}_{\scriptstyle x_3,x_4 \rightarrow y}}}}}} }\!\!\!
\bigtriangledown^{a}_{\!\!\! \mbox{}_{x_1}}\!\!\hspace{0.02ex}
\bigtriangledown^{b}_{\!\!\! \mbox{}_{x_2}}\!\!
\bigtriangledown^{c}_{\!\!\! \mbox{}_{x_3}}\!\!
\bigtriangledown^{d}_{\!\!\! \mbox{}_{x_4}}\!
{\rm T}\! \left(\hat{\Phi}_{n}(x_1)\hat{\Phi}_{n}(x_2)
\hat{\Phi}_{n}(x_3)\hat{\Phi}_{n}(x_4)  \right)\!,
\label{T star}
\ee
where ${\rm T}$ is the usual time ordering.
This ${\rm T}^{\displaystyle \ast}$ ``time ordering'' arises because
we have path integrals containing products of derivatives of the
field, which can be expressed as derivatives of the path
integrals which do not contain such derivatives. 
Notice, from the definitions 
(\ref{kernels}), that all the kernels which appear 
in expressions (\ref{derivatives}) are real and that 
$H_{\scriptscriptstyle \!{\rm A}_{\scriptstyle n}}^{abcd}$ is also
free of ultraviolet divergencies in the limit 
$n \!\rightarrow \!4$.

From (\ref{derivatives}) and (\ref{kernels}),  
it is clear that the imaginary part of the 
influence action, which does not contribute to the semiclassical 
Einstein equation (\ref{semiclassical eq in n}) 
because the expectation value of 
$\hat{T}_{n}^{ab}[g]$ is real, contains information on the
fluctuations of this operator. From (\ref{s-t expect value}) and
(\ref{derivatives}), taking into account that 
$S_{\rm IF}[g,g]=0$ and that 
$S_{\rm IF}[g^-,g^+]=
-S^{ {\displaystyle \ast}}_{\rm IF}[g^+,g^-]$, we can write the
expansion for the influence action 
$S_{\rm IF}[g\!+\!h^+,g\!+\!h^-]$ around a background
metric $g_{ab}$ in terms of the kernels (\ref{kernels}).
Taking into account that 
these kernels satisfy the symmetry relations
\be
H_{\scriptscriptstyle \!{\rm S}_{\scriptstyle n}}^{abcd}(x,y)=
H_{\scriptscriptstyle \!{\rm S}_{\scriptstyle n}}^{cdab}(y,x), 
\hspace{3 ex} 
H_{\scriptscriptstyle \!{\rm A}_{\scriptstyle n}}^{abcd}(x,y)=
-H_{\scriptscriptstyle \!{\rm A}_{\scriptstyle n}}^{cdab}(y,x), 
\hspace{3 ex}
K_n^{abcd}(x,y) = K_n^{cdab}(y,x),
\label{symmetries}
\ee
and introducing a new kernel 
\be
H_n^{abcd}(x,y)\equiv 
H_{\scriptscriptstyle \!{\rm S}_{\scriptstyle n}}^{abcd}(x,y)
+H_{\scriptscriptstyle \!{\rm A}_{\scriptstyle n}}^{abcd}(x,y),
\label{H}
\ee
this expansion can be finally written as
\bea
S_{\rm IF}[g\!+\!h^+,g+h^-]
&=& {1\over 2} \int\! d^nx\, \sqrt{- g(x)}\:
\langle \hat{T}_{n}^{ab}(x) \rangle [g] \,
\left[h_{ab}(x) \right]  \nn \\
&&
-\,{1\over 2} \int\! d^nx\, d^ny\, \sqrt{- g(x)}\sqrt{- g(y)}\,
\left[h_{ab}(x)\right]
\left(H_n^{abcd}[g](x,y)\!
+\!K_n^{abcd}[g](x,y) \right)
\left\{ h_{cd}(y) \right\}  \nn  \\
&&
+\,{i\over 2} \int\! d^nx\, d^ny\, \sqrt{- g(x)}\sqrt{- g(y)}\,
\left[h_{ab}(x) \right]
N_n^{abcd}[g](x,y)
\left[h_{cd}(y) \right]+0(h^3),
\label{expansion 2}
\eea 
where we have used the notation
\be
\left[h_{ab}\right] \equiv h^+_{ab}\!-\!h^-_{ab},
\hspace{5 ex}
\left\{ h_{ab}\right\} \equiv h^+_{ab}\!+\!h^-_{ab}.
\label{notation}
\ee

We are now in the position to carry out the formal derivation of the
semiclassical Einstein-Langevin equation. 
The procedure is well known
\cite{calzettahu,humatacz,husinha,cv96,lomb-mazz,cv97,ccv97,%
campos-hu,campos-hu2,calver98,caldeira,hu-matacz94,hu-paz-zhang2,%
greiner,matacz,morikawa,shaisultanov,gleiser}, 
it consists of deriving a new
``improved'' effective action using the 
the following identity:
\be
e^{-{1\over 2} \!\int\! d^nx\, d^ny \, \sqrt{- g(x)}\sqrt{- g(y)}\,
\left[h_{ab}(x) \right]\, N_n^{abcd}(x,y)\, \left[h_{cd}(y)\right] }=
\int\! {\cal D}[\xi_n]\: {\cal P}[\xi_n]\, e^{i \!\int\! d^nx \,
\sqrt{- g(x)}\,\xi_n^{ab}(x)\,\left[h_{ab}(x) \right] },
\label{Gaussian path integral}
\ee
where ${\cal P}[\xi_n]$ is the probability distribution
functional of a Gaussian stochastic tensor $\xi_n^{ab}$ 
characterized by the correlators
\be
\langle\xi_n^{ab}(x) \rangle_{c}\!= 0, 
\hspace{10ex} 
\langle\xi_n^{ab}(x)\xi_n^{cd}(y) \rangle_{c}\!=
N_n^{abcd}[g](x,y),
\label{correlators in n}
\ee
with $N_n^{abcd}$ given in (\ref{kernels}), 
and where
the path integration measure is assumed to be a scalar under
diffeomorphisms of $({\cal M},g_{ab})$. The above identity follows
from the identification of the right hand side of
(\ref{Gaussian path integral}) with the characteristic functional for 
the stochastic field $\xi_n^{ab}$.
In fact, by differentiation of this expression 
with respect to $\left[h_{ab}\right]$, it can be checked that this is
the characteristic functional of a stochastic field characterized by the
correlators (\ref{correlators in n}).
When $N_n^{abcd}(x,y)$ is strictly positive definite, the 
probability distribution functional for $\xi_n^{ab}$ is explicitly
given by  
\be
{\cal P}[\xi_n]=
 \frac{e^{-{1\over2}\!\int\! d^nx\, d^ny \, \sqrt{-g(x)}\sqrt{-g(y)}\,
 \xi_n^{ab}(x) \, N^{-1}_{n\,abcd}(x,y)\, \xi_n^{cd}(y)}}
 {\int\! {\cal D}\bigl[\bar{\xi}_n\bigr]\:
   e^{-{1\over2}\!\int\! d^nz\, d^nw \, \sqrt{-g(z)}\sqrt{-g(w)}\,
  \bar{\xi}_n^{ef}(z) \, N^{-1}_{n\,efgh}(z,w)\, \bar{\xi}_n^{gh}(w)}},
\label{gaussian probability}
\ee
where $N^{-1}_{n\,abcd}[g](x,y)$ is the inverse of
$N_n^{abcd}[g](x,y)$ defined by
\be
\int\! d^nz \, \sqrt{- g(z)}\, N_n^{abef}(x,z) N^{-1}_{n\,efcd}(z,y)=
{1\over2} \left(\delta^a_c \delta^b_d+\delta^a_d \delta^b_c \right)
{\delta^n(x\!-\!y) \over \sqrt{- g(x)}}.
\label{inverse}
\ee
Using the identity 
(\ref{Gaussian path integral}), we can write the modulus of the
influence functional in the approximation (\ref{expansion 2}) as 
\be
\bigl|\hspace{0.2ex}{\cal F}_{\rm IF}[g+h^+,g+h^-]
\hspace{0.2ex}\bigr|=
e^{-{\rm Im}\, S_{\rm IF}[g+h^+,g+h^-]}=
\left\langle e^{i \int\! d^nx \,
\sqrt{- g(x)}\,\xi_n^{ab}(x)\,\left[h_{ab}(x) \right] } 
\right\rangle_{\! c}
\label{infl funct modulus}
\ee
where $\langle \hspace{1.5ex} \rangle_c$ means statistical average
over the stochastic tensor $\xi_n^{ab}$.
Thus, the effect of the imaginary part of the influence action 
(\ref{expansion 2}) on the corresponding influence
functional is equivalent to the averaged effect of the stochastic
source $\xi_n^{ab}$ coupled linearly 
to the perturbations $h_{ab}^{\pm}$.
The influence functional, in the approximation 
(\ref{expansion 2}), can be written as a statistical average over 
$\xi_n^{ab}$:
\be
{\cal F}_{\rm IF}[g+h^+,g+h^-]= \left\langle 
e^{i {\cal A}^{\rm eff}_{\rm IF}[h^+,h^-;g;\xi_n]}
\right\rangle_{\! c},
\label{infl funt as average}
\ee
with
\be
{\cal A}^{\rm eff}_{\rm IF}[h^+,h^-;g;\xi_n] \equiv 
{\rm Re}\, S_{\rm IF}[g\!+\!h^+,g\!+\!h^-]+\!
\int\! d^nx \,
\sqrt{- g(x)}\,\xi_n^{ab}(x)\left[h_{ab}(x) \right]+0(h^3),
\label{eff influence action}
\ee
where ${\rm Re}\, S_{\rm IF}$ can be read from the expansion
(\ref{expansion 2}). 
Note that the stochastic term in this action contains the information
of the imaginary part of $S_{\rm IF}$. 
Introducing a new ``improved'' effective action 
\be
{\cal A}_{\rm eff}[h^+,h^-;g;\xi_n] \equiv S_{g}[g+h^+]-S_{g}[g+h^-]+
{\cal A}^{\rm eff}_{\rm IF}[h^+,h^-;g;\xi_n],
\label{stochastic eff action}
\ee
where $S_{g}[g+h^{\pm}]$ has to be expanded up to second order
in the perturbations $h_{ab}^{\pm}$,
the equation of motion for 
$h_{ab}$ can be derived as  
\be
\left. {1\over\sqrt{-\det (g\!+\!h)(x)}} \, 
\frac{\delta {\cal A}_{\rm eff}[h^+,h^-;g;\xi_n]}{\delta h^+_{ab}(x)} 
\right|_{h^+=h^-=h}=0.
\label{eq of motion}
\ee
From (\ref{perturb s-t expect value}), taking into account that
only the real part of the influence action contributes to the
expectation value of the stress-energy tensor, we get,
to linear order in $h_{ab}$, 
\be
{1\over 8 \pi G_{B}}\biggl(  G^{ab}[g\!+\!h]+ 
\Lambda_{B} \left(g^{ab}\!-\!h^{ab}\right) \biggr)-
\left({4\over 3}\, \alpha_{B} D^{ab}
+ 2 \beta_{B} B^{ab} \right)\![g \!+\! h] = \mu^{-(n-4)} 
\langle \hat{T}_{n}^{ab} \rangle [g\!+\!h]
+2 \mu^{-(n-4)} \xi_n^{ab}, 
\label{Einstein-Langevin eq in n} 
\ee
where 
$h^{ab}\!\equiv\! g^{ac}g^{bd}h_{cd}$, that is, 
$g^{ab}\!- h^{ab}\!+ 0(h^2)$ is
the inverse of the metric $g_{ab}\!+\!h_{ab}$. 
This last equation is the semiclassical Einstein-Langevin 
equation in dimensional regularization.
As we have pointed out in section \ref{sec:E-L}, 
the two-point correlation function of the stochastic source 
in this equation [see Eq.~(\ref{correlators in n})], 
given by the 
noise kernel defined in (\ref{kernels}), is
free of ultraviolet divergencies in the limit $n \!\rightarrow \!4$.
Therefore, in the Einstein-Langevin equation
(\ref{Einstein-Langevin eq in n}),
one can perform exactly the same renormalization procedure
as for the semiclassical Einstein equation 
(\ref{semiclassical eq in n}).
After this, 
Eq.~(\ref{Einstein-Langevin eq in n}) will yield the physical 
semiclassical Einstein-Langevin equation 
(\ref{Einstein-Langevin eq}).
The derivation presented in this paper clarifies the
physical meaning of the stochastic source formally introduced in the
effective action (\ref{eff influence action}) by the identification 
(\ref{infl funct modulus}), since it links its two-point correlation
function to the stress-energy fluctuations
by Eqs.~(\ref{correlators in n}) and (\ref{kernels}).

There is also a connection 
between the equations obtained by this formal functional 
method and the equations derived from the (in general, also formal) 
assumption that 
decoherence and classicalization of suitably coarse-grained
system variables is achieved
through the mechanism proposed by Gell-Mann 
and Hartle \cite{gell-mann-hartle} in the consistent histories
formulation of a quantum theory. This last approach allows to evaluate
the probability distribution associated to such decoherent variables,
given by the diagonal elements of a decoherence functional, and, under
some approximations, to derive effective quasiclassical equations of
motion for them. These effective equations of motion 
can be shown to 
coincide \cite{mv98_2} with the semiclassical equations for the
background and the 
Langevin-type equations for perturbations obtained from the above
functional method.
Taking this connection into account, 
we can also conclude that, if one formally assumes that the
Gell-Mann and Hartle mechanism works for the metric field, one is 
lead to the semiclassical Einstein equation and the semiclassical 
Einstein-Langevin equation for the background metric and for the
metric perturbations, respectively \cite{mv98_2}.

We end this subsection with some comments on the relation between the 
semiclassical Einstein-Langevin equation (\ref{Einstein-Langevin eq})
and the Langevin-type equations for stochastic metric perturbations
recently derived in the literature 
\cite{calzettahu,humatacz,husinha,cv96,lomb-mazz,cv97,ccv97,%
campos-hu,campos-hu2,calver98}.
In these previous derivations, one starts with the influence functional
(\ref{path integral}), with the state of the scalar field assumed to
be an ``in'' vacuum or 
an ``in'' thermal state, and computes explicitly the expansion for the
corresponding influence action around a specific metric background. 
One then applies the above formal method to
derive a Langevin equation for the perturbations to this background. 
However, most of these derivations start with a
``mini-superspace'' model and, thus, the metric perturbations are
assumed from the beginning to have a restrictive form. In those cases,
the derived Langevin equations do not correspond exactly to 
our equation, 
Eq.~(\ref{Einstein-Langevin eq}), but to a ``reduced''
version of this equation, in which only some components of the noise
kernel in Eq.~(\ref{correlators}) (or some particular combinations of
them) influence the dynamics of the metric perturbations. 
Only those
equations which have been derived starting from a completely general
form for the metric perturbations 
\cite{cv96,lomb-mazz,campos-hu,campos-hu2}
are actually particular cases of the 
semiclassical Einstein-Langevin equation 
(\ref{Einstein-Langevin eq}).
Note, however, that the stochastic equation derived in 
Refs.~\cite{campos-hu,campos-hu2} do not correspond exactly to
Eq.~(\ref{Einstein-Langevin eq}), since the background 
(Minkowski spacetime and a scalar field in a thermal state)
is not a solution of semiclassical gravity. 
In this case, for the reasons explained in Sec.~\ref{sec:E-L}, the
equation for the metric perturbations is not gauge invariant.

%%%%%%%%%%%%%%%%%%%%%%%%%%%%%%%%%%%%%%%%

\subsection{Explicit linear form 
of the Einstein-Langevin equation}
\label{subsec:explicit}

%%%%%%%%%%%%%%%%%%%%%%%%%%%%%%%%%%%%%%%%

We can write Eq.~(\ref{Einstein-Langevin eq in n}) in a more explicit
form by working out the expansion of 
$\langle \hat{T}_{n}^{ab}\rangle [g\!+\!h]$
up to linear order in the
perturbation $h_{ab}$. 
From Eq.~(\ref{perturb s-t expect value}), 
we see that this expansion can
be easily obtained from (\ref{expansion 2}).
Noting, from (\ref{kernels}), that  
\be
K_n^{abcd}[g](x,y)= -{1\over 4} \,
\langle \hat{T}_{n}^{ab}(x) \rangle [g] \,
{g^{cd}(x)\over\sqrt{- g(y)}}\, 
\delta^n(x\!-\!y)-{1\over 2}\,{1\over\sqrt{- g(y)}}
\left\langle \left.
\frac{\delta T^{ab}[g,\Phi_{n}](x)}{\delta g_{cd}(y)} 
\right|_{\Phi_{n}=\hat{\Phi}_{n}}\right\rangle \![g],
\label{K}
\ee
we get
\be
\langle \hat{T}_n^{ab}(x) \hspace{-0.1ex}\rangle 
[g\!+\!h] =
\langle \hat{T}_n^{ab}(x) \hspace{-0.1ex}\rangle 
[g] +
\langle 
\hat{T}_n^{{\scriptscriptstyle (1)}\hspace{0.1ex} ab}
[g;h](x) \hspace{-0.1ex} \rangle  [g] -
2 \!\int\! \hspace{-0.2ex} d^ny \,
\sqrt{- g(y)} \hspace{0.2ex}  H_n^{abcd}[g](x,y) \hspace{0.2ex} 
h_{cd}(y) + 0(h^2),
\label{s-t expect value expansion}
\ee
where the operator 
$\hat{T}_n^{{\scriptscriptstyle (1)}\hspace{0.1ex} ab}$ is defined
from the term of first order in the expansion of 
$T^{ab}[g+h,\Phi_{n}]$ as
\be
T^{ab}[g\!+\!h,\Phi_{n}]=T^{ab}[g,\Phi_{n}]+
T^{{\scriptscriptstyle (1)}\hspace{0.1ex} ab}[g,\Phi_{n};h]
+0(h^2),  \hspace{3.5 ex}
\hat{T}_n^{{\scriptscriptstyle (1)}\hspace{0.1ex} ab}
[g;h]\equiv
T^{{\scriptscriptstyle (1)}\hspace{0.1ex} ab}[g,\hat{\Phi}_{n}[g];h], 
\label{T(1)}
\ee
using, as always, a Weyl ordering prescription for the operators in the
last definition. Note that the third term on the right hand side of
Eq.~(\ref{s-t expect value expansion}) is a consequence of 
the dependence
on $h_{cd}$ of the field operator $\hat{\Phi}_{n}[g+h]$ and of the
density operator $\hat{\rho}[g+h]$.

Substituting (\ref{s-t expect value expansion}) into 
(\ref{Einstein-Langevin eq in n}), and taking into account that
$g_{ab}$ satisfies the semiclassical Einstein equation
(\ref{semiclassical eq in n}), we can write the Einstein-Langevin
equation (\ref{Einstein-Langevin eq in n}) as
\bea
&&{1\over 8 \pi G_{B}}\left(
G^{{\scriptscriptstyle (1)}\hspace{0.1ex} ab}
[g;h](x)-
\Lambda_{B}\, h^{ab}(x) \right) -
{4\over 3}\, \alpha_{B} D^{{\scriptscriptstyle (1)}\hspace{0.1ex} ab}
[g;h](x)
-2\beta_{B} B^{{\scriptscriptstyle (1)}\hspace{0.1ex} ab}
[g;h](x)   \nn \\
&&- \,\mu^{-(n-4)} \langle 
\hat{T}_n^{{\scriptscriptstyle (1)}\hspace{0.1ex} ab}
[g;h](x) \rangle [g]+
2 \!\int\! d^ny\, \sqrt{- g(y)}\,\mu^{-(n-4)} 
H_n^{abcd}[g](x,y)\, h_{cd}(y)
\hspace{-0.2ex}=
2 \mu^{-(n-4)} \xi_n^{ab}(x). 
\label{Einstein-Langevin eq 2} 
\eea 
In the last
equation we have used the superindex ${\scriptstyle (1)}$ to denote
the terms of first order in the 
expansion in $h_{ab}$ of the tensors $G^{ab}[g+h]$,
$D^{ab}[g+h]$ and $B^{ab}[g+h]$. Thus,
for instance, $G^{ab}[g+h]\!=\!G^{ab}[g]+
G^{{\scriptscriptstyle (1)}\hspace{0.1ex} ab}[g;h]+ 0(h^2)$. 
The explicit expressions for the tensors
$G^{{\scriptscriptstyle (1)}\hspace{0.1ex} ab}[g;h]$,
$D^{{\scriptscriptstyle (1)}\hspace{0.1ex} ab}[g;h]$ and
$B^{{\scriptscriptstyle (1)}\hspace{0.1ex} ab}[g;h]$ can be
found in the Appendix of Ref.~\cite{mv98_2}, 
and
$T^{{\scriptscriptstyle (1)}\hspace{0.1ex} ab}[g,\Phi_{n};h]$
is given in
Appendix \ref{sec:expansions of tensors}. 
From 
$T^{{\scriptscriptstyle (1)}\hspace{0.1ex} ab}[g,\Phi_{n};h]$, we can
write an explicit expression for the operator 
$\hat{T}_n^{{\scriptscriptstyle (1)}\hspace{0.1ex} ab}$.
In fact,
using the Klein-Gordon equation, and
expressions (\ref{regul s-t 2}) and (\ref{diff operator}) for the
stress-energy operator, we have
\be
\hat{T}_n^{{\scriptscriptstyle (1)}\hspace{0.1ex} ab}
[g;h]=\left({1\over 2}\, g^{ab}h_{cd}-\delta^a_c h^b_d-
\delta^b_c h^a_d  \right) \hat{T}_{n}^{cd}[g]
+{\cal F}^{ab}[g;h]\, \hat{\Phi}_{n}^2[g],  
\label{T(1) operator}
\ee
where ${\cal F}^{ab}[g;h]$ is the differential operator
\bea
{\cal F}^{ab} &\equiv& \left(\xi\!-\!{1\over 4}\right)\!\! 
\left(h^{ab}\!-\!{1\over 2}\, g^{ab} h^c_c \right)\! \Box+
{\xi \over 2} \left[ 
\bigtriangledown^{c}\! \bigtriangledown^{a}\! h^b_c+
\bigtriangledown^{c}\! \bigtriangledown^{b}\! h^a_c- 
\Box h^{ab}-
\bigtriangledown^{a}\! \bigtriangledown^{b}\!  h^c_c-
g^{ab}\! \bigtriangledown^{c}\! \bigtriangledown^{d} h_{cd}
\right.   \nn \\
&&+\left. g^{ab} \Box h^c_c 
+\left( \bigtriangledown^{a} h^b_c+
\bigtriangledown^{b} h^a_c-\bigtriangledown_{\! c} 
\hspace{0.2ex} h^{ab}-
2 g^{ab}\! \bigtriangledown^{d}\! h_{cd} +
g^{ab}\! \bigtriangledown_{\! c} \! h^d_d
\right)\! \bigtriangledown^{c}
-g^{ab} h_{cd} \bigtriangledown^{c}\! \bigtriangledown^{d} 
\right]. 
\label{diff operator F}
\eea
It is understood that indices are raised 
with the background inverse metric $g^{ab}$ and that all the
covariant derivatives are associated to the metric $g_{ab}$.
Substituting  expression (\ref{T(1) operator}) into 
Eq.~(\ref{Einstein-Langevin eq 2}), 
and using the semiclassical equation
(\ref{semiclassical eq in n}) to get an expression for 
$\mu^{-(n-4)}
\langle \hat{T}_{n}^{ab}\rangle [g]$, we can
finally write the semiclassical Einstein-Langevin equation in
dimensional regularization as
\bea
&&{1\over 8 \pi G_{B}}\Biggl[
G^{{\scriptscriptstyle (1)}\hspace{0.1ex} ab}\!-\!
{1\over 2}\, g^{ab} G^{cd} h_{cd}+ G^{ac} h^b_c+G^{bc} h^a_c+ 
\Lambda_{B} \left( h^{ab}\!-\!{1\over 2}\, g^{ab} h^c_c \right) 
\Biggr](x) 
   \nn \\
&&
- \, 
{4\over 3}\, \alpha_{B} \left( D^{{\scriptscriptstyle
(1)}\hspace{0.1ex} ab}
-{1\over 2}\, g^{ab} D^{cd} h_{cd}+ D^{ac} h^b_c+D^{bc} h^a_c
\right)\! (x)
-2\beta_{B}\left( B^{{\scriptscriptstyle (1)}\hspace{0.1ex} ab}\!-\!
{1\over 2}\, g^{ab} B^{cd} h_{cd}+ B^{ac} h^b_c+B^{bc} h^a_c 
\right)\! (x)   \nn \\
&&- \, \mu^{-(n-4)}\, {\cal F}^{ab}_x 
\langle \hat{\Phi}_{n}^2(x) \rangle [g]
+2 \!\int\! d^ny \, \sqrt{- g(y)}\, \mu^{-(n-4)} 
H_n^{abcd}[g](x,y)\, h_{cd}(y)
=2 \mu^{-(n-4)} \xi^{ab}_n(x),  
\label{Einstein-Langevin eq 3} 
\eea
where the tensors $G^{ab}$, $D^{ab}$ and
$B^{ab}$ are computed from the semiclassical metric $g_{ab}$,
and where we have omitted the functional dependence on $g_{ab}$ and
$h_{ab}$ in $G^{{\scriptscriptstyle (1)}\hspace{0.1ex} ab}$,
$D^{{\scriptscriptstyle (1)}\hspace{0.1ex} ab}$,
$B^{{\scriptscriptstyle (1)}\hspace{0.1ex} ab}$ and
${\cal F}^{ab}$ to simplify the notation. 
Notice that, in Eq.~(\ref{Einstein-Langevin eq 3}), 
all the ultraviolet divergencies in
the limit $n \!\rightarrow \!4$, which must be removed by
renormalization of the coupling constants, are in 
$\langle \hat{\Phi}_{n}^2(x) \rangle$ and the
symmetric part 
$H_{\scriptscriptstyle \!{\rm S}_{\scriptstyle n}}^{abcd}(x,y)$ of the
kernel  $H_n^{abcd}(x,y)$, whereas the
kernels $N_n^{abcd}(x,y)$ and
$H_{\scriptscriptstyle \!{\rm A}_{\scriptstyle n}}^{abcd}(x,y)$ are
free of ultraviolet divergencies. 
These two last kernels can be 
written in terms of 
$F_{n}^{abcd}[g](x,y) \equiv
\left\langle \hat{t}_n^{ab}(x)\,\hat{t}_n^{cd}(y)
  \right\rangle \![g]$
as
\be
N_n^{abcd}[g](x,y))=
{1\over 4}\,{\rm Re} \, F_{n}^{abcd}[g](x,y), 
\hspace{7ex}
H_{\scriptscriptstyle \!{\rm A}_{\scriptstyle n}}^{abcd}[g](x,y)=
{1\over 4}\,{\rm Im} \, F_{n}^{abcd}[g](x,y),   
\label{finite kernels}
\ee
where we have used that 
$2 \left\langle \hat{t}_n^{ab}(x)\, \hat{t}_n^{cd}(y)  \right\rangle=
\left\langle \left\{ \hat{t}_n^{ab}(x), \, \hat{t}_n^{cd}(y)
\right\}\right\rangle +
\left\langle \left[ \hat{t}_n^{ab}(x), 
\, \hat{t}_n^{cd}(y)\right]\right\rangle$, 
and the fact that the first term on the right hand side of this
identity is real, whereas the second one is pure imaginary.
Once we perform the renormalization procedure
in Eq.~(\ref{Einstein-Langevin eq 3}), setting
$n \!= \!4$ will yield the physical semiclassical
Einstein-Langevin equation. 
Note that, due to the presence of the kernel $H_n^{abcd}(x,y)$,
this equation will be usually non-local in the metric perturbation.

%%%%%%%%%%%%%%%%%%%%%%%%%%%%%%%%%%%%%%%%

\subsection{The kernels for a vacuum state}
\label{subsec:vacuum}

%%%%%%%%%%%%%%%%%%%%%%%%%%%%%%%%%%%%%%%%

We conclude this section by considering the case in which the 
expectation values that appear in the Einstein-Langevin equation 
(\ref{Einstein-Langevin eq 3}) [see Eqs.~(\ref{kernels})] 
are taken in a vacuum state $|0 \rangle$
(for a field quantized on $({\cal M},g_{ab})$ in the Heisenberg
picture), such as, for instance, an ``in'' vacuum.  
In this case we can go further and write these 
expectation values in terms of the Wightman and Feynman
functions, defined as
\be
G_n^+(x,y) \equiv \langle 0| \,
   \hat{\Phi}_{n}(x)  \hat{\Phi}_{n}(y) \,
   |0 \rangle [g],
\hspace{5 ex}
i G\!_{\scriptscriptstyle F_{\scriptstyle \hspace{0.1ex}  n}}
 \hspace{-0.2ex}(x,y) 
  \equiv \langle 0| \,
  {\rm T}\! \left( \hat{\Phi}_{n}(x)  \hat{\Phi}_{n}(y) \right)
  \hspace{-0.2ex}
  |0 \rangle [g].
\label{Wightman and Feynman functions}
\ee 
These expressions for the kernels in the Einstein-Langevin 
equation will be very useful for explicit 
calculations.
To simplify the notation, we omit the functional 
dependence on the semiclassical metric $g_{ab}$, which will be
understood in all the expressions below.

From (\ref{finite kernels}), we see that the kernels
$N_n^{abcd}(x,y)$ and 
$H_{\scriptscriptstyle \!{\rm A}_{\scriptstyle n}}^{abcd}(x,y)$ 
are the real and imaginary parts, respectively, of 
$F_{n}^{abcd}(x,y) \!=\!
\langle 0| \,  \hat{T}_n^{ab}(x)\, \hat{T}_n^{cd}(y)\, |0 \rangle
\!-\!
\langle 0| \,  \hat{T}_n^{ab}(x)\,|0 \rangle 
\langle 0| \,\hat{T}_n^{cd}(y)\, |0 \rangle$. Since, from
(\ref{regul s-t 2}), we can write the operator $\hat{T}_n^{ab}$ as a
sum of terms of the form $\left\{ {\cal A}_x \hat{\Phi}_{n}(x), 
\,{\cal B}_x \hat{\Phi}_{n}(x)\right\}$, where ${\cal A}_x$ and 
${\cal B}_x$ are some differential operators, we can express 
$F_{n}^{abcd}(x,y)$ in
terms of the Wightman function using 
\bea
&&\left\langle   
\left\{ {\cal A}_x \hat{\Phi}_{n}(x), 
{\cal B}_x \hat{\Phi}_{n}(x)\right\} \hspace{-0.4ex} 
\left\{ {\cal C}_y \hat{\Phi}_{n}(y), 
{\cal D}_y \hat{\Phi}_{n}(y)\right\} \hspace{-0.1ex}
\right\rangle
\!-\! \left\langle  
\left\{ {\cal A}_x \hat{\Phi}_{n}(x), 
{\cal B}_x \hat{\Phi}_{n}(x)\right\} \hspace{-0.1ex} \right\rangle
\left\langle
\left\{ \hspace{-0.1ex} {\cal C}_y \hat{\Phi}_{n}(y), 
{\cal D}_y \hat{\Phi}_{n}(y)\right\} \hspace{-0.1ex}
\right\rangle   \hspace{20ex}
    \nn   \\
&& \hspace{41ex}
= 4 \,{\cal A}_x {\cal C}_y G_n^+(x,y)\, 
{\cal B}_x{\cal D}_y G_n^+(x,y)+
4 \,{\cal A}_x {\cal D}_y G_n^+(x,y)\, 
{\cal B}_x{\cal C}_y G_n^+(x,y),
\label{Wightman expression}
\eea
where ${\cal C}_x$ and ${\cal D}_x$ are also
some differential operators and where the expectation values are
taken in the vacuum $|0 \rangle$. 
This identity 
can be easily proved using Wick's theorem or by 
writing the operator $\hat{\Phi}_{n}(x)$ in terms of the creation
and annihilation operators of the Fock representation corresponding to
the vacuum $|0 \rangle$. Using a Schwinger-DeWitt expansion for the
Wightman  function $G_n^+(x,y)$, one can actually see that the two
terms on the right hand side of the last expression are free of
ultraviolet divergencies in the limit $n\!\rightarrow \! 4$. 
Finally, we find 
\bea
F_{n}^{abcd}(x,y)
&=&\bigtriangledown^{a}_{\!\!\! \mbox{}_{x}}\!
 \bigtriangledown^{c}_{\!\!\! \mbox{}_{y}}\! G_n^+(x,y) 
 \bigtriangledown^{b}_{\!\!\! \mbox{}_{x}}\!
 \bigtriangledown^{d}_{\!\!\! \mbox{}_{y}} G_n^+(x,y)
+\bigtriangledown^{a}_{\!\!\! \mbox{}_{x}}\!
 \bigtriangledown^{d}_{\!\!\! \mbox{}_{y}}\! G_n^+(x,y) 
 \bigtriangledown^{b}_{\!\!\! \mbox{}_{x}}\!
 \bigtriangledown^{c}_{\!\!\! \mbox{}_{y}} G_n^+(x,y)
\hspace{32.5ex}
   \nn \\
&& 
+\, 2\, {\cal D}^{ab}_{\!\! \scriptscriptstyle x}  \bigl(
  \bigtriangledown^{c}_{\!\!\! \mbox{}_{y}} G_n^+(x,y)
  \bigtriangledown^{d}_{\!\!\! \mbox{}_{y}}\! G_n^+(x,y) \bigr)
+2\, {\cal D}^{cd}_{\!\! \scriptscriptstyle y} \bigl(
  \bigtriangledown^{a}_{\!\!\! \mbox{}_{x}} G_n^+(x,y)
  \bigtriangledown^{b}_{\!\!\! \mbox{}_{x}}\! G_n^+(x,y) \bigr)
+2\, {\cal D}^{ab}_{\!\! \scriptscriptstyle x} 
   {\cal D}^{cd}_{\!\! \scriptscriptstyle y}  \bigl(
 G_n^{+ 2}(x,y)  \bigr),  
\label{Wightman expression 2}
\eea
where ${\cal D}^{ab}_{\!\! \scriptscriptstyle x}$ is the differential
operator (\ref{diff operator}). From this expression and the relations
(\ref{finite kernels}), we get expressions for the kernels
$N_n^{abcd}(x,y)$ and 
$H_{\scriptscriptstyle \!{\rm A}_{\scriptstyle n}}^{abcd}(x,y)$ in
terms of the Wightman function $G_n^+(x,y)$.

The kernel 
$H_{\scriptscriptstyle \!{\rm S}_{\scriptstyle n}}^{abcd}(x,y)$,
defined in (\ref{kernels}), can be written in terms of the Feynman
function noting that, from Wick's theorem, 
\bea
&&{\rm Im} \, \Bigl\langle  {\rm T}^{\displaystyle \ast}\! \Bigl( 
\left\{ {\cal A}_x \hat{\Phi}_{n}(x), 
{\cal B}_x \hat{\Phi}_{n}(x)\right\} \hspace{-0.3ex} 
\left\{ {\cal C}_y \hat{\Phi}_{n}(y), 
{\cal D}_y \hat{\Phi}_{n}(y)\right\} \Bigr)
\Bigr\rangle
    \nn   \\
&&\hspace{20ex} =- 4 \, {\rm Im} \Bigl[ {\cal A}_x {\cal C}_y 
  G\!_{\scriptscriptstyle F_{\scriptstyle \hspace{0.1ex}  n}}
 \hspace{-0.2ex}(x,y)\, 
{\cal B}_x{\cal D}_y 
  G\!_{\scriptscriptstyle F_{\scriptstyle \hspace{0.1ex}  n}}
 \hspace{-0.2ex}(x,y)+
{\cal A}_x {\cal D}_y
  G\!_{\scriptscriptstyle F_{\scriptstyle \hspace{0.1ex}  n}}
 \hspace{-0.2ex}(x,y) \, 
{\cal B}_x{\cal C}_y 
  G\!_{\scriptscriptstyle F_{\scriptstyle \hspace{0.1ex}  n}}
 \hspace{-0.2ex}(x,y) \Bigr],
\label{Feynman expression}
\eea
where, again, 
${\cal A}_x$, ${\cal B}_x$, ${\cal C}_x$ and ${\cal D}_x$ are
real differential operators and the expectation value is
in the vacuum $|0 \rangle$. The kernel 
$H_{\scriptscriptstyle \!{\rm S}_{\scriptstyle n}}^{abcd}(x,y)$ is then 
obtained by adding up the contribution of all the differential 
operators which appear in the product $T^{ab}(x) T^{cd}(y)$, where 
$T^{ab}$ is the functional (\ref{class s-t}). 
After a long calculation, we get
\bea
H_{\scriptscriptstyle \!{\rm S}_{\scriptstyle n}}^{abcd}(x,y)=
- {1 \over 4} \, {\rm Im} \Bigl[ &&
 \bigtriangledown^{a}_{\!\!\! \mbox{}_{x}}\!
 \bigtriangledown^{c}_{\!\!\! \mbox{}_{y}}\!
     G\!_{\scriptscriptstyle F_{\scriptstyle \hspace{0.1ex}  n}}
 \hspace{-0.2ex}(x,y)
 \bigtriangledown^{b}_{\!\!\! \mbox{}_{x}}\!
 \bigtriangledown^{d}_{\!\!\! \mbox{}_{y}}
     G\!_{\scriptscriptstyle F_{\scriptstyle \hspace{0.1ex}  n}}
 \hspace{-0.2ex}(x,y)
+\bigtriangledown^{a}_{\!\!\! \mbox{}_{x}}\!
 \bigtriangledown^{d}_{\!\!\! \mbox{}_{y}}\! 
     G\!_{\scriptscriptstyle F_{\scriptstyle \hspace{0.1ex}  n}}
 \hspace{-0.2ex}(x,y)
 \bigtriangledown^{b}_{\!\!\! \mbox{}_{x}}\!
 \bigtriangledown^{c}_{\!\!\! \mbox{}_{y}}
     G\!_{\scriptscriptstyle F_{\scriptstyle \hspace{0.1ex}  n}}
 \hspace{-0.2ex}(x,y)   \nn \\
&& 
-\,g^{ab}(x) \bigtriangledown^{e}_{\!\!\! \mbox{}_{x}}\!
 \bigtriangledown^{c}_{\!\!\! \mbox{}_{y}}
     G\!_{\scriptscriptstyle F_{\scriptstyle \hspace{0.1ex}  n}}
 \hspace{-0.2ex}(x,y)
 \bigtriangledown_{\!\!e}^{ \mbox{}_{x} }\!
 \bigtriangledown^{d}_{\!\!\! \mbox{}_{y}}
     G\!_{\scriptscriptstyle F_{\scriptstyle \hspace{0.1ex}  n}}
 \hspace{-0.2ex}(x,y)
-g^{cd}(y) \bigtriangledown^{a}_{\!\!\! \mbox{}_{x}}\!
 \bigtriangledown^{e}_{\!\!\! \mbox{}_{y}}
     G\!_{\scriptscriptstyle F_{\scriptstyle \hspace{0.1ex}  n}}
 \hspace{-0.2ex}(x,y)
 \bigtriangledown^{b}_{\!\!\! \mbox{}_{x}}
 \bigtriangledown_{\!\!e}^{ \mbox{}_{y} }
     G\!_{\scriptscriptstyle F_{\scriptstyle \hspace{0.1ex}  n}}
 \hspace{-0.2ex}(x,y)    \nn  \\
&& 
+\,{1 \over 2}\, g^{ab}(x) g^{cd}(y) 
 \bigtriangledown^{e}_{\!\!\! \mbox{}_{x}}\!
 \bigtriangledown^{f}_{\!\!\! \mbox{}_{y}}
     G\!_{\scriptscriptstyle F_{\scriptstyle \hspace{0.1ex}  n}}
 \hspace{-0.2ex}(x,y)
 \bigtriangledown_{\!\!e}^{ \mbox{}_{x} }\!
 \bigtriangledown_{\!\!f}^{ \mbox{}_{y} }
     G\!_{\scriptscriptstyle F_{\scriptstyle \hspace{0.1ex}  n}}
 \hspace{-0.2ex}(x,y)
+{\cal K}^{ab}_{\! \scriptscriptstyle x}  \bigl(
 2 \hspace{-0.2ex} \bigtriangledown^{c}_{\!\!\! \mbox{}_{y}}\!
   G\!_{\scriptscriptstyle F_{\scriptstyle \hspace{0.1ex}  n}}
   \hspace{-0.2ex}(x,y)
 \bigtriangledown^{d}_{\!\!\! \mbox{}_{y}}\!
   G\!_{\scriptscriptstyle F_{\scriptstyle \hspace{0.1ex}  n}}
   \hspace{-0.2ex}(x,y)
     \nn   \\
&&   
  -\, g^{cd}(y) \bigtriangledown^{e}_{\!\!\! \mbox{}_{y}}\!
   G\!_{\scriptscriptstyle F_{\scriptstyle \hspace{0.1ex}  n}}
   \hspace{-0.2ex}(x,y)
\bigtriangledown_{\!\!e}^{ \mbox{}_{y} }\!
     G\!_{\scriptscriptstyle F_{\scriptstyle \hspace{0.1ex}  n}}
     \hspace{-0.2ex}(x,y) \bigr)
+{\cal K}^{cd}_{\! \scriptscriptstyle y}  \bigl(
 2 \hspace{-0.2ex} \bigtriangledown^{a}_{\!\!\! \mbox{}_{x}}\!
   G\!_{\scriptscriptstyle F_{\scriptstyle \hspace{0.1ex}  n}}
   \hspace{-0.2ex}(x,y)
 \bigtriangledown^{b}_{\!\!\! \mbox{}_{x}}\!
   G\!_{\scriptscriptstyle F_{\scriptstyle \hspace{0.1ex}  n}}
   \hspace{-0.2ex}(x,y)
     \nn   \\
&&  
-\, g^{ab}(x) \bigtriangledown^{e}_{\!\!\! \mbox{}_{x}}\!
   G\!_{\scriptscriptstyle F_{\scriptstyle \hspace{0.1ex}  n}}
   \hspace{-0.2ex}(x,y)
\bigtriangledown_{\!\!e}^{ \mbox{}_{x} }\!
     G\!_{\scriptscriptstyle F_{\scriptstyle \hspace{0.1ex}  n}}
     \hspace{-0.2ex}(x,y) \bigr) 
+2\, {\cal K}^{ab}_{\! \scriptscriptstyle x}
   {\cal K}^{cd}_{\! \scriptscriptstyle y}  \bigl(
   G\!_{\scriptscriptstyle F_{\scriptstyle \hspace{0.1ex}  n}}^{\;\: 2}
   \hspace{-0.2ex}(x,y)  \bigr) \Bigr],
\label{Feynman expression 2}
\eea 
where ${\cal K}^{ab}_{\! \scriptscriptstyle x}$ is the differential
operator
\be
{\cal K}^{ab}_{\! \scriptscriptstyle x} \equiv 
\xi \left( g^{ab}(x) \Box_{\! \scriptscriptstyle x}
  -\bigtriangledown^{a}_{\!\!\! \mbox{}_{x}}\!
   \bigtriangledown^{b}_{\!\!\! \mbox{}_{x}}+\, G^{ab}(x) \right)
-{1 \over 2}\, m^2 g^{ab}(x).
\label{diff operator K}
\ee
An alternative expression for 
$H_{\scriptscriptstyle \!{\rm S}_{\scriptstyle n}}^{abcd}(x,y)$,
which is more similar to expression (\ref{Wightman expression 2}),
can be obtained taking into account that 
$G\!_{\scriptscriptstyle F_{\scriptstyle \hspace{0.1ex}  n}}
 \hspace{-0.2ex}(x,y)$ is a Green function of the Klein-Gordon
equation in $n$ spacetime dimensions, which satisfies
\be
\left( \Box_{\! \scriptscriptstyle x} -m^2- \xi R(x) \right)
G\!_{\scriptscriptstyle F_{\scriptstyle \hspace{0.1ex}  n}}
     \hspace{-0.2ex}(x,y)={\delta^n(x\!-\!y) \over \sqrt{- g(x)}},
\label{Green function eq}
\ee
and using that in dimensional regularization 
$\left[\delta^n(x\!-\!y) \right]^2=0$. 
Finally, note that, in the vacuum 
$|0 \rangle$, the term
$\langle \hat{\Phi}_{n}^2 (x) \rangle$ in
equation (\ref{Einstein-Langevin eq 3}) can also be written as
$\langle \hat{\Phi}_{n}^2(x) \rangle=
i G\!_{\scriptscriptstyle F_{\scriptstyle \hspace{0.1ex}  n}}
      \hspace{-0.2ex}(x,x)=G_n^+(x,x)$.

It is worth noting that, when the points $x$ and $y$ are spacelike
separated, $\hat{\Phi}_{n}(x)$ and $\hat{\Phi}_{n}(y)$ commute and,
thus, $G_n^+(x,y) \!=\! 
i G\!_{\scriptscriptstyle F_{\scriptstyle \hspace{0.1ex}  n}}
 \hspace{-0.2ex}(x,y) \!=\! 
(1/2) \langle 0| \,
\{ \hat{\Phi}_{n}(x) , \hat{\Phi}_{n}(y) \} \,  |0 \rangle$, which is
real. Hence, from the above expressions, we have that 
$H_{\scriptscriptstyle \!{\rm A}_{\scriptstyle n}}^{abcd}(x,y) \!=\!
H_{\scriptscriptstyle \!{\rm S}_{\scriptstyle n}}^{abcd}(x,y) \!=\!
0$. 
This fact is not surprising since, from the causality 
of the expectation value of the stress-energy operator, we know that  
the non-local dependence on the metric perturbation in the
Einstein-Langevin equation must be causal.

%%%%%%%%%%%%%%%%%%%%%%%%%%%%%%%%%%%%%%%%

\section{Fluctuations 
in stationary and conformally stationary backgrounds}
\label{sec:stationary}

%%%%%%%%%%%%%%%%%%%%%%%%%%%%%%%%%%%%%%%%

In this section, we derive a number of results concerning the
stochastic semiclassical theory of gravity for two classes of
background solutions of semiclassical gravity. The first class
consists of a stationary spacetime and a scalar field in thermal
equilibrium or in its vacuum state. 
In the second class, the spacetime is
conformally stationary, the scalar field is massless and conformally
coupled, and its state is the conformal vacuum or a
thermal state built on the conformal vacuum.
In subsections \ref{subsec:f-d in stationary} and 
\ref{subsec:f-d in conformally stationary}, we identify a kernel in
the corresponding Einstein-Langevin equations which is related to the
noise kernel by a fluctuation-dissipation relation. 
In subsection \ref{subsec:particle creation}, we study the creation of
particles by stochastic metric perturbations and see that 
this phenomenon can be related to the vacuum noise kernel.  
We show that the mean value
of created particles is enhanced by the presence of metric
fluctuations with respect to the same quantity 
in the ``perturbed'' semiclassical spacetime
$({\cal M},g_{ab} \!+\! \langle h_{ab} \rangle_c)$.

Let us assume that the semiclassical 
spacetime $({\cal M},g_{ab})$ is stationary, {\it i.e.}, that 
it possesses a global timelike Killing vector field 
$\zeta^a$, ${\cal \pounds}_{\mbox{}_{\! \zeta}}g_{ab}=0$, where 
${\cal \pounds}_{\mbox{}_{\! \zeta}}$ is the Lie 
derivative with respect to $\zeta^a$. 
Writing the Killing vector as $\zeta^a=(\partial / \partial t)^a$,
this spacetime can be 
foliated by a family of Cauchy hypersurfaces $\Sigma_{t}$, labeled by
the Killing time $t$, 
so we can give coordinates $(t,{\bf x})$ to each
spacetime point, where ${\bf x}$ are the space
coordinates on each of these hypersurfaces. 
Using this foliation, we can construct a
Hamiltonian operator  
$\hat{H}[g]$ in the way described in Appendix \ref{sec:Hamiltonian}. 
This is a time independent, {\it i.e.},
independent of the Cauchy hypersurface $\Sigma_{t}$, Hamiltonian
operator, so it represents the Hamiltonian operator
in both the Heisenberg and the Schr\"{o}dinger pictures. 
In this case, there is a natural Fock representation based
on a decomposition of the field operator $\hat{\Phi}_{n}[g]$ in 
a complete set of modes of positive frequencies $\omega_{k}$ 
with respect to $\zeta^a$, and their complex 
conjugates.\footnote{In some cases, 
additional restrictions may be necessary to
avoid infrared divergencies, such as that the scalar field is massive,
$m^2 \!\neq\! 0$, or that the norm of the Killing vector is not
arbitrarily small \cite{wald94,kay}.}
This defines a natural Fock space, the many-particle states of which
are eigenstates of the Hamiltonian $\hat{H}[g]$. Thus, 
the notion of particles is
physically well defined in this spacetime
\cite{wald94,kay,mostepanenko}. 
The Hamiltonian operator in this Fock representation, renormalized by
normal ordering, is given by
$\hat{H}[g]\!=\!\sum_{k} \omega_{k} \,\hat{a}_k^{\dag} \hat{a}_k$, 
where $\hat{a}_k^{\dag}$ and $\hat{a}_k$ 
are the creation and annihilation
operators on the Fock space. Here, 
the summation must be understood as
representing either a sum over a set of discrete indices or an
integral with some suitable measure (or a combination of these two
possibilities). 
The time-evolution operator corresponding to this Hamiltonian operator
is then given by 
$\hat{\cal U}[g](t,t^{\prime})
\! \equiv \exp \hspace{0.2ex}\bigl(-i \hat{H}[g] \,
(t\!-\!t^{\prime})\bigr)$.

In this section, even if we sometimes write $t_i$ or $t_f$, we shall
always consider these initial and final times in the limit 
$t_i \!\rightarrow \! -\infty$ and $t_f \!\rightarrow \! +\infty$
(we assume that such limits can be taken).

%%%%%%%%%%%%%%%%%%%%%%%%%%%%%%%%%%%%%%%%

\subsection{The fluctuation-dissipation 
relation in a stationary background}
\label{subsec:f-d in stationary}

%%%%%%%%%%%%%%%%%%%%%%%%%%%%%%%%%%%%%%%%

For a real scalar field quantized on the stationary spacetime 
$({\cal M},g_{ab})$, we can define a state of thermal equilibrium 
at temperature $T$.
This state is described in the Heisenberg picture
by the density operator of the grand canonical ensemble: 
\be
\hat{\rho}[g]={e^{-\beta \hat{H}[g]} \over 
{\rm Tr} \! \left(e^{-\beta \hat{H}[g]} \right) },
\label{thermal state}
\ee 
where $\beta \equiv 1/k_B T$ and $k_B$ is Boltzmann's constant
(there are no chemical potential terms 
because we deal with a real scalar field).
This kind of thermal states for fields in stationary curved
backgrounds was first considered in 
Refs.~\cite{dowker77,gibbons78}. 
Since the density operator (\ref{thermal state})
commutes with the time-evolution operator
$\hat{\cal U}[g](t,t^{\prime})$, the corresponding 
initial density operator in the Schr\"{o}dinger picture is simply
$\hat{\rho}^{\rm \scriptscriptstyle S}(t_i)\!=\!\hat{\rho}[g]$.

Given any pair of operators in the Heisenberg picture, 
$\hat{P}[g](x)$ and $\hat{Q}[g](x)$, 
the expectation value 
$\langle \hat{P}(x) \hat{Q}(x^{\prime })
\rangle_{\mbox{}_{\mbox{}_{\! \scriptscriptstyle T}}}[g]$
depends on $t$ and $t^{\prime}$ only through the difference
$t-t^{\prime}$, since
\be
\langle \hat{P}(x) \hat{Q}(x^{\prime })
\rangle_{\mbox{}_{\mbox{}_{\! \scriptscriptstyle T}}}=
{\rm Tr} \! \left[ \hat{\rho}\, \hat{P}(x) 
\hat{Q}(x^{\prime }) \right]=
{\rm Tr} \! \left[ \hat{\rho}\, 
\hat{P}^{\rm \scriptscriptstyle S}({\bf x}) 
e^{-i \hat{H}(t-t^{\prime})} 
\hat{Q}^{\rm \scriptscriptstyle S}({\bf x^{\prime}}) 
e^{i \hat{H}(t-t^{\prime})} \right],
\ee 
where $\hat{P}^{\rm \scriptscriptstyle S}({\bf x})$ and
$\hat{Q}^{\rm \scriptscriptstyle S}({\bf x})$ are 
the operators in the Schr\"{o}dinger picture corresponding to 
$\hat{P}(x)$ and $\hat{Q}(x)$, respectively, and we use
$\langle \hspace{1.5ex} \rangle
_{\mbox{}_{\mbox{}_{\! \scriptscriptstyle T}}}$ to denote an
expectation value in the state described by
(\ref{thermal state}).
In particular, with the choice 
$\hat{\rho}^{\rm \scriptscriptstyle S}(t_i)\!=\!\hat{\rho}[g]$,
the kernels 
$N_{n }^{abcd}[g](x,x^{\prime })$,
$H_{\scriptscriptstyle \!{\rm S}_{\scriptstyle n
 }}^{abcd}[g](x,x^{\prime })$ and
$H_{\scriptscriptstyle \!{\rm A}_{\scriptstyle n
 }}^{abcd}[g](x,x^{\prime })$
depend on the time coordinates
as a function of $t-t^{\prime}$.
Therefore, we
can introduce Fourier transforms in the time coordinate as
\be
K(x,x^{\prime})= \int^{\infty}_{-\infty} {d\omega \over 2 \pi}\, 
e^{-i \omega (t-t^{\prime})}\, 
\tilde{K}(\omega;{\bf x},{\bf x^{\prime}}),
\label{Fourier transform}
\ee
where $K(x,x^{\prime})$ is any function which depends on time only 
through $t\!-\!t^{\prime}$.

As it is shown in Appendix \ref{sec:Hamiltonian}, Wick's theorem can
be generalized for thermal $N$-point functions, defined as expectation
values
of products of the field operator in the state described by 
(\ref{thermal state}). It is then easy to see that the expressions
found in subsection \ref{subsec:vacuum} also hold for the kernels
$N_{n }^{abcd}[g](x,x^{\prime })$,
$H_{\scriptscriptstyle \!{\rm S}_{\scriptstyle n
 }}^{abcd}[g](x,x^{\prime })$ and
$H_{\scriptscriptstyle \!{\rm A}_{\scriptstyle n
 }}^{abcd}[g](x,x^{\prime })$ at finite $T$ if
we replace the Wightman and Feynman functions 
(\ref{Wightman and Feynman functions}) by the analogous thermal
expectation values.

In this case, a simple relationship (in the form of a 
fluctuation-dissipation relation) exists 
between the kernels 
$N_{n }^{abcd}[g](x,x^{\prime })$ and
$H_{\scriptscriptstyle \!{\rm A}_{\scriptstyle n
 }}^{abcd}[g](x,x^{\prime })$. 
In fact, from (\ref{finite kernels}), we can write these kernels
as
\be
8\, N_{n }^{abcd}(x,x^{\prime })=
F_{n}^{abcd}(x,x^{\prime })+
F_{n}^{cdab}(x^{\prime },x),
\hspace{5ex}
8 i\, H_{\scriptscriptstyle \!{\rm A}_{\scriptstyle n
 }}^{abcd}(x,x^{\prime })=
F_{n}^{abcd}(x,x^{\prime })-
F_{n}^{cdab}(x^{\prime },x),
\label{rel}
\ee
where we omit the functional dependence on $g_{ab}$.
In terms of the Fourier transforms 
(\ref{Fourier transform}), these relations are
\bea
8\, \tilde{N}_{n }^{abcd}
(\omega;{\bf x},{\bf x^{\prime}})&=&
\tilde{F}_{n}^{abcd}
(\omega;{\bf x},{\bf x^{\prime}})+
\tilde{F}_{n}^{cdab}
(-\omega;{\bf x^{\prime}},{\bf x}),    \nn  \\
8 i\, \tilde{H}_{\scriptscriptstyle \!{\rm A}_{\scriptstyle n
 }}^{abcd}
(\omega;{\bf x},{\bf x^{\prime}})&=&
\tilde{F}_{n}^{abcd}
(\omega;{\bf x},{\bf x^{\prime}})+
\tilde{F}_{n}^{cdab}
(-\omega;{\bf x^{\prime}},{\bf x}).
\label{relations}
\eea
By analytically continuing $t$ to complex
values in $F_{n}^{abcd}(x,x^{\prime })$, 
one can derive a symmetry relation for this bi-tensor which involves
different values of this complex time. Taking into account that the 
time evolution of the operator $\hat{t}_n^{ab}$ is given in this
stationary case by
$\hat{t}_n^{ab}(t+\Delta t,{\bf x})=
e^{i \hat{H} \Delta t}\, \hat{t}_n^{ab}(t,{\bf x}) \, 
e^{-i \hat{H} \Delta t}$, and using the
cyclic property of the trace, we get 
$F_{n}^{abcd}(t,{\bf x};
t^{\prime },{\bf x}^{\prime })= F_{n}^{cdab}
(t^{\prime },{\bf x}^{\prime };t+i \beta,{\bf x})$,
or, equivalently, in terms of its Fourier transform, 
\be
\tilde{F}_{n}^{abcd}
(\omega;{\bf x},{\bf x^{\prime}})= e^{\beta \omega} 
\tilde{F}_{n}^{cdab}
(-\omega;{\bf x^{\prime}},{\bf x}).
\ee
This relation is known as the Kubo-Martin-Schwinger relation
\cite{kubo,martin}.
From this last expression and (\ref{relations}), we obtain the
following simple relation between 
$\tilde{N}_{n }^{abcd}$ and 
$\tilde{H}_{\scriptscriptstyle \!{\rm A}_{\scriptstyle n
 }}^{abcd}$:
\be
\tilde{H}_{\scriptscriptstyle \!{\rm A}_{\scriptstyle n
 }}^{abcd}
(\omega;{\bf x},{\bf x^{\prime}})=
-i \, \tanh\! \left( {\beta \omega \over 2} \right)
\tilde{N}_{n }^{abcd}
(\omega;{\bf x},{\bf x^{\prime}}),
\label{f-d relation}
\ee
which can also be written as
\be
H_{\scriptscriptstyle \!{\rm A}_{\scriptstyle n
 }}^{abcd}
(t,{\bf x};t^{\prime },{\bf x}^{\prime })=
\int^{\infty}_{-\infty} dt^{\prime \prime}\, 
K_{\scriptscriptstyle {\rm FD} }
(t-t^{\prime \prime})\,
N_{n }^{abcd}
(t^{\prime \prime},{\bf x};t^{\prime },{\bf x}^{\prime }),
\label{f-d relation 2}
\ee
with
\be
K_{\scriptscriptstyle {\rm FD} }(t)
\equiv -\int^{\infty}_{0} {d\omega \over  \pi}\,  
\sin (\omega t)\hspace{0.2ex}
 \tanh\! \left( {\beta \omega \over 2} \right)=
-k_B T \: {\rm P} \hspace{-0.3ex} \left[\hspace{0.2ex}
{\rm cosech}\, (\pi k_B T \hspace{0.2ex} t) \right],
\ee
where ${\rm P}$ denotes a Cauchy principal value distribution.

Since, as we have pointed out above, the kernels 
$H_{\scriptscriptstyle \!{\rm A}_{\scriptstyle n
 }}^{abcd}$ and 
$N_{n }^{abcd}$ are free of ultraviolet
divergencies in the limit 
$n\!\rightarrow \! 4$, we can define
\be
H_{\scriptscriptstyle \!{\rm A} }^{abcd}
(x,x^{\prime })
\equiv \lim_{n \rightarrow 4}\, \mu^{-2 (n-4)}
H_{\scriptscriptstyle \!{\rm A}_{\scriptstyle n
 }}^{abcd}(x,x^{\prime }),
\hspace{5ex}
N ^{abcd}(x,x^{\prime }) \equiv
\lim_{n \rightarrow 4}\, \mu^{-2 (n-4)} 
N_{n }^{abcd}(x,x^{\prime }),
\label{physical kernels}
\ee
which are the kernels that appear in the physical semiclassical
Einstein-Langevin equation, Eq.~(\ref{Einstein-Langevin eq}), after
performing the renormalization procedure in 
Eq.~(\ref{Einstein-Langevin eq 3}).
These physical kernels will also satisfy the relation 
(\ref{f-d relation 2}) or, equivalently, their Fourier transforms will
satisfy (\ref{f-d relation}). 
These results are independent of the 
regularization method used.

The relation (\ref{f-d relation}) can be written in 
an alternative way. 
Introducing a new kernel (this is actually a family of
kernels) defined by 
$\tilde{H}_{\scriptscriptstyle \!{\rm A}_{\scriptstyle n
 }}^{abcd}
(\omega;{\bf x},{\bf x^{\prime}}) \equiv -i \omega \, 
\tilde{\gamma}_{n }^{abcd}
(\omega;{\bf x},{\bf x^{\prime}})$, that is,
$H_{\scriptscriptstyle \!{\rm A}_{\scriptstyle n
 }}^{abcd}(x,x^{\prime})
= \partial 
\gamma_{n }^{abcd}(x,x^{\prime})
/ \partial t$, Eq.~(\ref{f-d relation}) yields
\be
\tilde{N}_{n }^{abcd}
(\omega;{\bf x},{\bf x^{\prime}}) = 
\omega \: 
{\rm cotanh }\! \left( {\beta \omega \over 2} \right)
\tilde{\gamma}_{n }^{abcd}
(\omega;{\bf x},{\bf x^{\prime}}),
\label{f-d relation 3}
\ee
or, equivalently,
\be
N_{n }^{abcd}
(t,{\bf x};t^{\prime },{\bf x}^{\prime })=
\int^{\infty}_{-\infty} dt^{\prime \prime}\, 
J_{\scriptscriptstyle {\rm FD} }
(t-t^{\prime \prime})\,
\gamma_{n }^{abcd}
(t^{\prime \prime},{\bf x};t^{\prime },{\bf x}^{\prime }),
\label{f-d relation 4}
\ee
where
\be
J_{\scriptscriptstyle {\rm FD} }(t)
\equiv \int^{\infty}_{0} {d\omega \over  \pi}\,  
\cos (\omega t)\: \omega \: 
{\rm cotanh }\! \left( {\beta \omega \over 2} \right).
\ee
This integral gives a distribution which is singular at
$t \!=\! 0$ and for $t \! \neq \! 0$ reduces to 
$J_{\scriptscriptstyle {\rm FD} }(t)
\! = \! - \pi \, [k_B T \: {\rm cosech}( \pi k_B T \, t)]^2$.

The relations (\ref{f-d relation}) or  
(\ref{f-d relation 2}) [or the equivalent forms (\ref{f-d relation 3})
or (\ref{f-d relation 4})] have the same form as the
fluctuation-dissipation relations which appear in quite general 
models of quantum mechanics 
\cite{landau,kubo,grabert,schwinger61,weber,kubo85}. 
The derivation of these relations is usually done in the framework of  
linear response theory, in which one considers the response of a 
quantum system, which is
initially at thermal equilibrium, when an external classical
time-dependent linear perturbation is ``switched on.'' 
When evaluating the change in the
expectation value of the relevant operator (the operator which
couples to the perturbation) induced by the presence 
of the perturbation,  
a dissipative term can be identified as the term
which changes the sign under a time reversal 
transformation in the perturbation. 
This term is characterized
by a kernel called the dissipation kernel. It can be shown that the
dissipation kernel is related to the fluctuations in equilibrium (in the
absence of the perturbation) of the relevant operator
by a relation which is exactly the same as (\ref{f-d relation 2}) or
(\ref{f-d relation}). This is the fluctuation-dissipation relation.
Using this linear response theory approach, the same
fluctuation-dissipation relation has also been derived for
some models of quantum many-body systems \cite{martin,kubo85} or quantum
fields \cite{weber,jackiw} coupled to external classical fields.

This fluctuation-dissipation relation appears also in the context of
quantum Brownian motion (or ``semiclassical'' Brownian motion),  
in which one is interested in 
the dynamics of a macroscopic particle
in interaction with a heat bath environment, usually modelized by an
infinite set of quantum harmonic oscillators. 
In these models, when the variable representing the center of mass
position of the macroscopic particle decoheres, 
it can be effectively described as a
classical stochastic variable.
The equation of motion for this stochastic variable is 
a linear Langevin equation with a Gaussian 
stochastic source. The classical variable introduced in linear
response theory can be envisaged as the position of the Brownian
particle, but now this variable becomes a dynamical stochastic
variable.  
The dissipative term in this Langevin equation is the responsible for
the irreversible dynamics of the Brownian particle. 
This term contains a
dissipation kernel which is related to the correlator
of the stochastic source by the relations (\ref{f-d relation 2}) or 
(\ref{f-d relation}) \cite{caldeira,hu-paz-zhang}. 
This is again the fluctuation-dissipation relation. 
There are also some models 
in which a purely quantum description of the Brownian particle is
considered \cite{lindenberg,kac}.
The dynamics of this particle is then described by a quantum
operator in the Heisenberg picture. By elimination of all the
environment degrees of freedom in the equation of motion for this
operator, one finds a quantum Langevin equation with quantum
fluctuating and dissipative terms. These terms are again related by a
fluctuation-dissipation relation of the form (\ref{f-d relation 2}) or 
(\ref{f-d relation}).

These analogies allow us to identify the equivalent relations 
(\ref{f-d relation 2}) and (\ref{f-d relation}), 
and the analogous relations
for the physical kernels (\ref{physical kernels}), as the 
fluctuation-dissipation relation in our context. Because of this
relation, the kernel 
$H_{\scriptscriptstyle \!{\rm A} }^{abcd}
(x,x^{\prime })$
shall be called the dissipation kernel.
The same fluctuation-dissipation relation was derived by Mottola
\cite{mottola} in the context of quantum field theory in curved
spacetime using the linear response theory approach. This author
considered the case in which the background spacetime is static, but
his result is easily generalized to a stationary background.
In this paper, we have derived the
same relation in the context of a Langevin equation for stochastic
metric perturbations, which would presumably describe the
effective dynamics of gravitational fluctuations after a process of
decoherence. 
For the particular case of a massless scalar field in a Minkowski
background, this fluctuation-dissipation relation was
derived in Refs.~\cite{campos-hu,campos-hu2} 
from an explicit evaluation of the kernels.

It is clear that the kernel
$N ^{abcd}(x,x^{\prime })$ describes
fluctuations in exactly the same sense as the quantum-mechanical
models described above. In fact, as it was pointed out by Mottola 
\cite{mottola} from the point of view of linear
response theory, it gives the fluctuations in equilibrium of the
stress-energy operator. Alternatively, as we have shown in the previous
sections, it gives the two-point correlation
function of the Gaussian stochastic source in the semiclassical
Einstein-Langevin equation.
However, the term containing the ``dissipation'' kernel
$H_{\scriptscriptstyle \!{\rm A} }^{abcd}
(x,x^{\prime })$
in the Einstein-Langevin equation does not generally 
change sign under a time-reversal transformation in the metric
perturbations.

\subsubsection{Zero temperature limit}
\label{subsub:zero T}

A state of the scalar field which is of special interest is that
described by $\hat{\rho}^{\rm \scriptscriptstyle S}(t_i)\!=\!
\hat{\rho}[g]\! =\! |0 \rangle \langle 0|$, where $|0 \rangle$ is the 
vacuum state. This vacuum
state can be obtained as the zero temperature limit,
$T\!\rightarrow \! 0$, of the previous thermal state.
The fluctuation-dissipation relation for this state is
easily obtained by setting $T=0$ in expression (\ref{f-d relation})
or (\ref{f-d relation 2}). We find 
$\tilde{H}_{\scriptscriptstyle \!{\rm A}_{\scriptstyle n
 }}^{abcd}
(\omega;{\bf x},{\bf x^{\prime}})=
-i \, {\rm sign}\, \omega \,
\tilde{N}_{n }^{abcd}
(\omega;{\bf x},{\bf x^{\prime}})$, 
or, equivalently, it has the form (\ref{f-d relation 2}),
with
\be
K_{\scriptscriptstyle {\rm FD} }(t)
= -i \int^{\infty}_{-\infty} {d\omega \over 2 \pi}\,  
e^{-i \omega t} \,{\rm sign}\, \omega =
-{1 \over \pi}\, 
{\rm P}\!\hspace{-0.1ex} \left( {1 \over t} \right).
\ee
This fluctuation-dissipation relation in the alternative form
(\ref{f-d relation 3}) reads
$\tilde{N}_{n }^{abcd}
(\omega;{\bf x},{\bf x^{\prime}}) = 
\omega \: 
{\rm sign}\, \omega \:
\tilde{\gamma}_{n }^{abcd}
(\omega;{\bf x},{\bf x^{\prime}})$,
or, it has the form (\ref{f-d relation 4}), with
\be
J_{\scriptscriptstyle {\rm FD} }(t)
= \int^{\infty}_{0} {d\omega \over  \pi}\,  
\cos (\omega t)\: \omega = 
- {1 \over \pi}\, 
{\cal P}\!f \!\hspace{-0.1ex} \left( {1 \over t^2} \right),
\ee
where ${\cal P}\!f  ( 1/ t^2 )$
denotes a Hadamard finite part distribution, which is related to
${\rm P}( 1/t )$ by 
${\cal P}\!f ( 1/ t^2 ) = - (d/dt) \hspace{0.2ex} {\rm P}( 1/t )$ 
(the definitions of these distributions can be found in 
Refs.~\cite{schwartz}).

\subsubsection{High temperature limit}

Let us now consider the high temperature limit. This limit can only be
performed when there exists a cutoff frequency $\Omega$, such that 
$\tilde{N}_{n }^{abcd}
(\omega;{\bf x},{\bf x^{\prime}})$ vanishes for 
$\omega\! >\!\Omega$ (by (\ref{f-d relation}), 
$\tilde{H}_{\scriptscriptstyle \!{\rm A}_{\scriptstyle n
 }}^{abcd}(\omega;{\bf x},{\bf x^{\prime}})$
will also vanish for these values of $\omega$). Such a cutoff
frequency is usually related to a characteristic cutoff frequency of
the environment degrees of freedom. The high temperature limit
corresponds to the limit in which $k_B T \!\gg \! \hbar \Omega$. 
In this limit (keeping only the leading order contributions), we 
expect that thermal fluctuations dominate over quantum fluctuations. 
To study this limit, it is
convenient to restore the dependence in $\hbar$ in the previous
results. For this, one has to multiply the constants
$\alpha_{B}$ and $\beta_{B}$ by 
$\hbar$ and the kernel $H_n^{abcd}$ by $1/ \hbar$ in the
Einstein-Langevin equation (\ref{Einstein-Langevin eq 3}), 
and change
the combination $\beta \omega$ by  $\hbar\beta \omega$ in the previous
expressions. In this limit, we can approximate 
$\tanh ( \hbar\beta \omega / 2 ) \simeq \hbar\beta \omega / 2$, 
and the fluctuation-dissipation relation
reduces to
\be
{1 \over\hbar} \, \tilde{H}_{\scriptscriptstyle \!{\rm A}
_{\scriptstyle n }}^{abcd}
(\omega;{\bf x},{\bf x^{\prime}})=
-i \:  {\omega \over 2 k_B T} \,
\tilde{N}_{n }^{abcd}
(\omega;{\bf x},{\bf x^{\prime}}),
\label{classical f-d relation}
\ee
or, equivalently, since in this case 
$K_{\scriptscriptstyle {\rm FD} }(t)
\simeq (\hbar / 2 k_B T) \, (d / dt)\, \delta (t)$,
\be
{1 \over\hbar} \, H_{\scriptscriptstyle \!{\rm A}_{\scriptstyle n
 }}^{abcd}
(t,{\bf x};t^{\prime },{\bf x}^{\prime })=
{1 \over 2 k_B T}\,  {\partial \over \partial t}\, 
N_{n }^{abcd}
(t,{\bf x};t^{\prime },{\bf x}^{\prime }).
\label{classical f-d relation 2}
\ee
Note that 
$(1 /\hbar) \, H_{\scriptscriptstyle \!{\rm A}_{\scriptstyle n
 }}^{abcd}$ is the kernel that appears in the 
Einstein-Langevin equation (\ref{Einstein-Langevin eq 3}) when one 
writes the dependence in $\hbar$ explicitly. This relation has the
same form as the classical Green-Kubo fluctuation-dissipation
relation which appears either in a classical theory of linear response
\cite{green,kubo} 
or in a classical theory of Brownian motion \cite{kac,zwanzig}.
Notice, from (\ref{classical f-d relation 2}), that in this high
temperature limit we can simply take 
$\gamma_{n }^{abcd}(x,x^{\prime})
= (\hbar / 2 k_B T) \, 
N_{n }^{abcd}(x,x^{\prime})$.

%%%%%%%%%%%%%%%%%%%%%%%%%%%%%%%%%%%%%%%%

\subsection{The fluctuation-dissipation relation 
for conformal fields in a conformally stationary background}
\label{subsec:f-d in conformally stationary}

%%%%%%%%%%%%%%%%%%%%%%%%%%%%%%%%%%%%%%%%

In the case of a massless conformally coupled scalar field ($m\!=\!0$
and $\xi\!=\!1/6$) and a conformally stationary solution of
semiclassical gravity [for instance, a Robertson-Walker (RW) 
spacetime], the fluctuation-dissipation relation derived in the previous
subsection can be generalized
when the state of the field in the background solution  
is the conformal vacuum or a thermal state built on the conformal
vacuum. 
In this case, the action (\ref{scalar field action}) for
the scalar field is conformally invariant. It is convenient to 
preserve this conformal invariance when working in dimensional
regularization. This can be done by changing in all the previous 
expressions which involve dimensional regularization the parameter
$\xi$ by the function $\xi(n)\!\equiv \!(n-2)/[4(n-1)]$ and, of
course, taking $m\!=\!0$. In this way, the dimensional regularized 
stress-tensor operator (\ref{regul s-t}) is traceless.
Let $({\cal M},\overline{g}_{ab})$ be a $n$ dimensional
conformally stationary spacetime, that is, a spacetime with a global
timelike conformal Killing vector field
$\zeta^a$: 
${\cal \pounds}_{\mbox{}_{\! \zeta}}\overline{g}_{ab}
=(2/n) \, \bbderiv^{c}\hspace{-0.5ex}\zeta_{c}
\:\overline{g}_{ab}$, where $\deriv_{\!a}$ 
is the covariant derivative associated to $\overline{g}_{ab}$.
This means that the metric $\overline{g}_{ab}$ is conformally related
to a stationary metric $g_{ab}$: 
$\overline{g}_{ab}(x)=e^{2 \varpi(x)} g_{ab}(x)$, where 
$\varpi(x)$ is a scalar function. 
As previously, writing $\zeta^a\!=\!(\partial / \partial t)^a$,
the semiclassical spacetime can be foliated by Cauchy hypersurfaces 
$\Sigma_{t}$ and coordinates $(t,{\bf x})$ can be assigned to the
spacetime points.

There is a ``natural'' Fock representation based
on a decomposition of the field operator
$\hat{\Phi}_{n}[\bar{g}]$ in terms
of a complete set of modes 
$\{ \bar{u}_{k_{\mbox{}_{\scriptstyle n}}}\hspace{-0.4ex}(x) \}$, 
solution of the Klein-Gordon equation
with metric $\overline{g}_{ab}$, 
of the form 
$\bar{u}_{k_{\mbox{}_{\scriptstyle n}}}\hspace{-0.4ex}(x)=
e^{-(n-2)\varpi(x)/2} \hspace{0.2ex}
u_{k_{\mbox{}_{\scriptstyle n}}}\hspace{-0.4ex}(x)$,
where 
$\{ u_{k_{\mbox{}_{\scriptstyle n}}}\hspace{-0.4ex}(x) \}$
is a complete set of mode solutions 
of the Klein-Gordon equation in $( {\cal M}, g_{ab})$
which are of positive frequencies $\omega_{k}$ 
with respect to $\zeta^a$.
Hence, in this sense, we can write the field operator as
$\hat{\Phi}_{n}[\bar{g}]= 
e^{-(n-2)\varpi/2} \hspace{0.2ex} \hat{\Phi}_{n}[g]$,
where $\hat{\Phi}_{n}[g]$ is the 
field operator in the stationary spacetime 
$( {\cal M}, g_{ab})$.
Assuming that no infrared divergencies are present, so that 
this quantum field theory construction is well defined, 
the conformal vacuum $|0 \rangle$ is defined as the vacuum 
state of the Fock space corresponding to this representation.
If $\hat{a}_k^{\dag}$ and $\hat{a}_k$ are the creation and annihilation
operators on this Fock space, this state satisfies 
$\hat{a}_k |0 \rangle\!=\!0$. 
As shown in Appendix \ref{sec:Hamiltonian}, in this case we can
construct a conserved energy operator which can be identified with the
Hamiltonian of a field quantized on $( {\cal M}, g_{ab})$:
$\hat{E}[\bar{g}] \!=\!
\hat{H}[g]\!=\!
\sum_{k} \omega_{k} \,\hat{a}_k^{\dag} \hat{a}_k$. 
This energy operator, however, is not a
time-evolution generator for the field operator 
$\hat{\Phi}_{n}[\bar{g}]$, it  
generates the time-evolution of the conformally related operator 
$\hat{\Phi}_{n}[g]$.
The many-particle states of the Fock space built on the conformal
vacuum are eigenstates of this energy operator.

From this energy operator, a state of thermal equilibrium
for a conformal scalar field quantized on  
$({\cal M},\overline{g}_{ab})$ can be defined using
the density operator (\ref{thermal state}).
Thermal equilibrium states defined in this way 
were first proposed by Gibbons and Perry
\cite{gibbons78}. These authors were inspired in a result by
Israel \cite{israel72} in the framework of 
relativistic kinetic theory, who found that thermal equilibrium
distribution functions can be defined for massless particles
in conformally stationary spacetimes.
A number of applications have been developed in the literature to 
study finite-temperature effects of quantum conformal fields 
in RW universes \cite{cooke77}
and in two-dimensional spacetimes \cite{cramer98}.

Let us begin with a solution of the semiclassical
Einstein equation (\ref{semiclassical eq in n})
consisting of a quantum conformal scalar field in   
a conformally stationary spacetime 
$({\cal M},\overline{g}_{ab})$, in the thermal state 
(\ref{thermal state}). 
Taking into account that the
action (\ref{scalar action}) with $m\!=\!0$ and 
$\xi\!=\!\xi(n)$ satisfies 
$S_m[\bar{g},\bar{\Phi}_{n}]\!=\!S_m[g,\Phi_{n}]$, where 
$\bar{\Phi}_{n} \!\equiv\! e^{-(n-2)\varpi/2} \Phi_{n}$, 
it is easy to see that 
$\hat{T}_{n}^{ab}[\bar{g}] =e^{-(n+2)\varpi} \hat{T}_{n}^{ab}[g]$. 
Therefore, the kernels evaluated in the thermal state at a 
temperature $T$ can be related to the corresponding kernels 
for the stationary background
$({\cal M},g_{ab})$. For the noise kernel, we have
\be
N_{n }^{abcd}[\bar{g}](x,x^{\prime })=
e^{-(n+2)\varpi(x)}\, e^{-(n+2)\varpi(x^{\prime })}\,
N_{n }^{abcd}[g](x,x^{\prime }), 
\label{conformal relations for kernels} 
\ee
and the same relation holds for the kernels
$H_{\scriptscriptstyle \!{\rm S}_{\scriptstyle n
 }}^{abcd}$ and
$H_{\scriptscriptstyle \!{\rm A}_{\scriptstyle n
 }}^{abcd}$.
Since the kernels $N_{n }^{abcd}[g]$ and 
$H_{\scriptscriptstyle \!{\rm A}_{\scriptstyle n
 }}^{abcd}[g]$ satisfy the relation
(\ref{f-d relation 2}) [or, equivalently, (\ref{f-d relation})], 
this leads to a fluctuation-dissipation relation 
between the kernels 
$N_{n }^{abcd}[\bar{g}]$ and 
$H_{\scriptscriptstyle \!{\rm A}_{\scriptstyle n
 }}^{abcd}[\bar{g}]$. The same relation holds
for the physical kernels obtained by taking the limit 
$n\!\rightarrow \! 4$ as in Eq.~(\ref{physical kernels}).
For the conformal vacuum state, which corresponds to 
$T\!=\!0$, the fluctuation-dissipation relation follows directly 
from the result of subsection \ref{subsub:zero T}.
In the particular case of a spatially flat
RW solution of semiclassical gravity, 
this conformal vacuum fluctuation-dissipation relation 
was obtained before in Ref.~\cite{cv96} 
after an explicit calculation of the corresponding
kernels. 
The same relation was derived in Ref.~\cite{husinha}
in the framework of a ``reduced'' version of the Einstein-Langevin
equation inspired in a Bianchi-I type ``mini-superspace'' model.

%%%%%%%%%%%%%%%%%%%%%%%%%%%%%%%%%%%%%%%%

\subsection{Particle creation}
\label{subsec:particle creation}

%%%%%%%%%%%%%%%%%%%%%%%%%%%%%%%%%%%%%%%%

Let us now return to the case in which $({\cal M},g_{ab})$ is
stationary, the scalar field has arbitrary mass $m$ and arbitrary
coupling parameter $\xi$, and consider the stochastic perturbation
$h_{ab}$.
Note that 
$({\cal M},g_{ab}\!+\!h_{ab})$ can be viewed as representing an
ensemble of spacetimes distributed according to some
probability distribution functional. We are in fact considering a
scalar field quantized on each of these spacetimes, described by the
operator $\hat{\Phi}[g\!+\!h]$, and the family of states
of the field, described by $\hat{\rho}[g\!+\!h]$.

Let $h_{ab}[\xi]$ be a solution to the
semiclassical Einstein-Langevin equation, 
Eq.~(\ref{Einstein-Langevin eq}), whose moments vanish for times
$t<t_{\rm \scriptscriptstyle I}$ or, at least, they vanish 
``asymptotically'' in the remote past 
($t \! \rightarrow \! - \infty$). This means that there is a 
``remote past epoch'' ($t<t_{\rm \scriptscriptstyle I}$ or 
$t \! \rightarrow \! - \infty$) in which $h_{ab}$ behaves
deterministically as a zero tensor. 
In that case, if we take 
$\hat{\rho}^{\rm \scriptscriptstyle S}(t_i)\!=\! 
|0 \rangle \langle 0|$, where $|0 \rangle$ is the vacuum of the
natural Fock space for the field quantized on $({\cal M},g_{ab})$,
and we consider the limit $t_i \! \rightarrow \! - \infty$,
we have $\hat{\rho}[g+h]\!=\! |0,{\rm in} \rangle
\langle 0,{\rm in}|$, where 
$|0,{\rm in} \rangle$ 
represents the family of 
``in'' vacua for the field quantized on 
$({\cal M},g_{ab}\!+\!h_{ab})$.
Treating $h_{ab}$ as a classical ``external'' perturbation, one could
construct a Hamiltonian operator $\hat{H}[g+h](t)$
in the Heisenberg picture for which 
$|0,{\rm in} \rangle$ would be the ground state
in the ``remote past epoch.'' 
However, at later times, due to the presence of the
perturbation $h_{ab}$, this 
``in'' vacuum state will generally 
not be the ground state of the Hamiltonian.
One then says that ``particles'' are created in the ``in''
vacuum.

Physically meaningful many particle ``out'' states, 
in particular, an ``out'' vacuum 
$|0,{\rm out} \rangle$
for the scalar field in each of the spacetimes 
$({\cal M},g_{ab}\!+\!h_{ab})$ can be defined if there is also a 
``far future epoch'' for which $h_{ab}$ vanishes (in the same
statistical sense as above), either in an exact way for 
$t > t_{\rm \scriptscriptstyle F}$ or ``asymptotically'' for 
$t \! \rightarrow \! + \infty$. When this is the case, 
the vacuum persistence amplitude 
$\langle 0,{\rm out}|0,{\rm in}\rangle [g\!+\!h]
\equiv e^{i W[g+h]}$ is given by the
following path integral:
\be
e^{i W[g+h]}=
\int \! {\cal D}[\Phi_n]\, 
\langle 0,{\rm out}|\Phi_n(t_2),t_2 \rangle [g\!+\!h]\;
\langle \Phi_n(t_1),t_1 |0,{\rm in}\rangle [g\!+\!h] \;
e^{i S_{m}[g+h,\Phi_n] },
\label{vacuum persistence amplitude}
\ee
where 
$|\varphi_n,t_{1}\rangle$ and $|\varphi_n,t_{2}\rangle$ 
denote, respectively, eigenstates of the field
operator $\hat{\Phi}_n[g\!+\!h](t,{\bf x})$ at some arbitrary times 
$t\!=\!t_1$ and $t\!=\!t_2$, where $t_2>t_1$, with
eigenvalue $\varphi_n({\bf x})$, and where the integration domain for
the action is between $t_1$ and $t_2$. The wave functionals 
$\langle \varphi_n,t_{1}|0,{\rm in}\rangle$ and 
$\langle \varphi_n,t_{2}|0,{\rm out}\rangle$ have in general a
dependence on the metric, which we have indicated 
in (\ref{vacuum persistence amplitude}).
In the limit 
$t_1 \!\rightarrow \! -\infty$ and
$t_2 \!\rightarrow \! +\infty$, these wave functionals 
do not depend on the perturbation $h_{ab}$. 
The total
probability of particle creation is given by \cite{cespedes}
\be
P[h;g] = 2 \lim_{n \rightarrow 4}\, {\rm Im}\, W[g\!+\!h].
\label{probability of particle creation}
\ee
One can show that ${\rm Im}\, W$ is free of ultraviolet
divergencies in the limit $n \!\rightarrow\! 4$, and that it is always
positive or zero, so that the probability $P$ is well defined by this
expression.

As we have done in the previous section for the influence action, we
can now expand the action $W[g+h]$ in the perturbation $h_{ab}$. 
In order to do so, one has to evaluate the functional derivatives
of $W[g+h]$ in the background metric $g_{ab}$. 
Using (\ref{vacuum persistence amplitude}), these derivatives
can be related to ``in-out'' matrix elements of operators
in the background. Since $g_{ab}$ is stationary, 
the ``in'' and ``out'' vacua in the background 
must be identified with the natural vacuum $|0 \rangle$. Therefore,
these background ``in-out'' matrix elements become expectation values
in the state $|0 \rangle$. It is then easy to see that 
the expansion of $W[g+h]$ in the metric perturbation $h_{ab}$ is equal
to that of $S_{\rm IF}[g+h^+,g+h^-]$ with
$h^+_{ab}=h_{ab}$ and 
$h^-_{ab}=0$, and taking the expectation values in 
$|0 \rangle$. In particular, from 
the imaginary part of this expansion 
[see Eq.~(\ref{expansion 2})], we get
\be
P[h;g]=\int\! d^4x\, d^4y \, \sqrt{- g(x)}\sqrt{- g(y)}\: h_{ab}(x)\, 
N^{abcd}[g](x,y)\,
h_{cd}(y)+0(h^3),
\label{stochastic probability} 
\ee
where $N^{abcd}$ is the zero temperature
physical noise kernel defined in (\ref{physical kernels}).
This physical noise kernel is 
related to the lowest order 
quantum stress-energy fluctuations in vacuum 
by (\ref{noise}). 
Note that the higher order corrections 
in (\ref{stochastic probability}) would contain higher order
vacuum stress-energy fluctuations.
Eq.~(\ref{stochastic probability}) is a generalization of an
expression derived by Sexl and Urbantke \cite{sexl} for the total
probability of particle creation by  
metric perturbations around Minkowski spacetime.

Eq.~(\ref{stochastic probability}) gives also the expectation value of
the number operator for ``out'' particles in the ``in'' vacuum,
computed to lowest order in the metric perturbation. 
In order to show this, let us expand the scalar field 
action as the action in the stationary 
background plus interaction terms (the terms containing the metric 
perturbation). 
The interaction term to lowest order in $h_{ab}$ is
$S^{(1)}\!=\! \int \! d^nx \, {\cal L}_n^{(1)}[\Phi_{n},h;g]$, with
${\cal L}_n^{(1)} \!=\! 
(1/2) \, \sqrt{- g} \:  T^{ab}[g,\Phi_{n}] \hspace{0.2ex} h_{ab}$. 
In order to construct the $S$-matrix operator, we need the interaction
Hamiltonian density operator in the interaction picture.
Note that the field and canonical momentum operators in the
interaction picture can be identified with the operators 
$\hat{\Phi}_{n}[g]$ and $\hat{\Pi}_{n}[g]$, respectively.   
Following Appendix \ref{sec:Hamiltonian}, we can obtain 
the canonical Hamiltonian density for the metric 
$g_{ab}\!+\!h_{ab}$ and 
work out the interaction term to first order 
in the metric perturbation. 
Although in this case the interaction Lagrangian density depends on
the derivatives of the field, we find that, to first order in 
$h_{ab}$, the interaction Hamiltonian density operator in the
interaction picture is given by 
$- {\cal L}_n^{(1)}[\hat{\Phi}_{n}[g],h;g]$. 
Hence, to first order in the metric perturbation, the 
$S$-matrix operator is given by 
$\hat{S} = 1 \!+\! \hat{S}^{(1)} \!+\! O(h^2)$, where
\be
\hat{S}^{(1)} = {i \over 2} \int \! d^nx \,
\sqrt{- g} \:  \hat{T}_n^{ab}[g] \, h_{ab}.
\label{S matrix}
\ee 
The expectation value of the ``out'' particle number operator, 
$\hat{N}^{\rm out}$, 
in the ``in'' vacuum (in the Heisenberg picture) is given by
$N[h;g] \equiv \langle 0,{\rm in}| \hspace{0.2ex} 
\hat{N}^{\rm out} \hspace{0.2ex} |0,{\rm in} \rangle
= \langle 0 | \hat{S}^{\dag} \hat{N} \hat{S} |0 \rangle$, 
where $\hat{N}$ is the particle number operator in the background
$\hat{N} \!\equiv \! \sum_{k} \hat{a}_k^{\dag} \hat{a}_k$.
To lowest order, we have 
\be
N[h;g] = \sum_{k,p} 
\left| \langle 1_k, 1_p | \hat{S}^{(1)} |0 \rangle \right|^2 
+ O(h^3),
\label{number of particles} 
\ee
where $| 1_k, 1_p \rangle$ is the two-particle state
$| 1_k, 1_p \rangle \equiv 
\hat{a}_k^{\dag} \hat{a}_p^{\dag} \, |0 \rangle$. 
Clearly, since $\hat{S}^{(1)}$ is quadratic in the field operator, 
at this order $N$ can also be written as
$N/2 = 
\sum_n \langle 0| \hat{S}^{(1)\, \dag} |n \rangle
\langle n| \hat{S}^{(1)} |0 \rangle
-\langle 0| \hat{S}^{(1)\, \dag} |0 \rangle
\langle 0| \hat{S}^{(1)} |0 \rangle+ O(h^3)$, where $\{ |n \rangle \}$
represents the complete orthonormal basis of the Fock space.
Using (\ref{S matrix}), this last expression can be written in terms
of the vacuum noise kernel 
$N_{n }^{abcd}[g](x,y)$ [see (\ref{kernels})].
Taking the limit $n \!\rightarrow \! 4$, we see that
the expression for one half of 
the number of created particles $N[h;g]/2$ to lowest order in the metric
perturbation coincides with that for $P[h;g]$ in 
Eq.~(\ref{stochastic probability}).

The energy of the created particles, defined as
$E[h;g] \equiv \langle 0,{\rm in}| \hspace{0.2ex} 
\sum_k \omega_k \hspace{0.2ex}
\hat{N}_k^{\rm out} \hspace{0.2ex} |0,{\rm in} \rangle
= \langle 0 | \hat{S}^{\dag} \hat{H} \hat{S} |0 \rangle$, 
where $\hat{N}_k^{\rm out}$ is the ``out'' number operator in the
$k$ mode and $\hat{H}\!=\!
\sum_{k} \omega_{k} \,\hat{a}_k^{\dag} \hat{a}_k$ 
is the Hamiltonian operator
in the background, is similarly given by
\be
E[h;g] = {1 \over 2} \sum_{k,p}  (\omega_k + \omega_p)
\left| \langle 1_k, 1_p | \hat{S}^{(1)} |0 \rangle \right|^2 
+ O(h^3).
\label{energy of particles}
\ee
Comparison of (\ref{energy of particles})
with (\ref{number of particles}) and 
(\ref{stochastic probability}), suggests that it may be possible 
in some cases to write 
this last expression (in the limit  
$n \!\rightarrow \! 4$) in terms of the
Fourier transform of the vacuum noise kernel.

As an example, let us consider 
the case when $({\cal M},g_{ab})$ is
$({\rm I\hspace{-0.4 ex}R}^{4},\eta_{ab})$
\cite{mv98,paperII}, which is the trivial solution of semiclassical
gravity. Working in a global
inertial coordinate system $\{x^\mu \}$, in this case
the kernels depend only on the difference $(x\!-\!y)^\mu$ and, thus,
we can define their Fourier transforms as
$K(x \!-\! y) \equiv (2\pi)^{-4} \!
\int \! d^4 p  \,
e^{i p \cdot (x-y)}\, \tilde{K}(p)$, where 
$p \hspace{-0.2ex} \cdot\! x \equiv \eta_{\mu \nu} p^\mu x^\nu$. 
Introducing the Fourier
transform of $h_{ab}(x)$ in a similar way
[note that $\tilde{h}_{ab}(-p) \!=\! 
\tilde{h}_{ab}^{\displaystyle \ast}(p)$], 
(\ref{stochastic probability}) can be written as 
\be
P[h;\eta] = \int \! {d^4 p \over (2\pi)^4 } \,
\tilde{N}^{abcd}(p) \, 
\tilde{h}_{ab}^{\displaystyle \ast}(p) \, 
\tilde{h}_{cd}(p) + O(h^3).
\label{prob in Minkowski}
\ee
On the other hand, the energy of the created particles is given by
\cite{frieman}
\be
E[h;\eta] = 2 \int \! {d^4 p \over (2\pi)^4 } \:
p^0 \, \theta(p^0) \,
\tilde{N}^{abcd}(p) \, 
\tilde{h}_{ab}^{\displaystyle \ast}(p) \, 
\tilde{h}_{cd}(p) + O(h^3).
\ee
The vacuum noise and dissipation kernels for a Minkowski background 
can be written in terms of two pairs of scalar kernels,
$N_r(x \!-\! y)$ and $D_r(x \!-\! y)$, respectively,
with $r \!=\! 1,2$ \cite{paperII} (see also Ref.~\cite{mv98} for a
particular case in which $N_2\!=\!D_2\!=\!0$).
Each pair of kernels $(N_r, D_r)$ satisfies the
fluctuation-dissipation relation found in 
subsection \ref{subsub:zero T}.
One finds \cite{cespedes,frieman} that
\be
\tilde{N}^{abcd}(p) \, 
\tilde{h}_{ab}^{\displaystyle \ast}(p) \, 
\tilde{h}_{cd}(p) = 
\tilde{C}^{{\scriptscriptstyle (1)}}_{abcd}(p)\,
\tilde{C}^{{\scriptscriptstyle (1)} 
{\displaystyle \hspace{0.1ex}\ast \hspace{0.1ex}}
abcd}(p) \, \tilde{N}_1(p)+
\left| \hspace{0.1ex} 
\tilde{R}^{{\scriptscriptstyle (1)}}(p)
\hspace{0.1ex}\right|^2 \tilde{N}_2(p),
\label{relation in Minkowski}
\ee
where $\tilde{C}^{{\scriptscriptstyle (1)}}_{abcd}(p)$,
$\tilde{R}^{{\scriptscriptstyle (1)}}(p)$ and $\tilde{N}_r(p)$
are, respectively, the
Fourier transforms of the linearized Weyl tensor, the scalar
curvature and the
kernels $N_r(x \!-\! y)$, $r \!=\! 1,2$;
$\tilde{N}_r(p)$ depend only on 
$p^2 \equiv \eta_{\mu \nu} p^\mu p^\nu$. It is then easy to see, using
the fluctuation-dissipation relation, that
\be
E[h;\eta] = i \int \! {d^4 p \over (2\pi)^4 } \:
p^0 \left[ \tilde{C}^{{\scriptscriptstyle (1)}}_{abcd}(p)\,
\tilde{C}^{{\scriptscriptstyle (1)} 
{\displaystyle \hspace{0.1ex}\ast \hspace{0.1ex}}
abcd}(p) \, \tilde{D}_1(p)+
\left| \hspace{0.1ex} 
\tilde{R}^{{\scriptscriptstyle (1)}}(p)
\hspace{0.1ex}\right|^2 \tilde{D}_2(p) \right]
+O(h^3).
\label{energy of particles in Minkowski}
\ee
Hence, in the case of a Minkowski background, the energy
of the created particles can be expressed in terms of the dissipation
kernels $D_1$ and $D_2$ for the Minkowskian vacuum.
It is not clear, however, that, for other
stationary backgrounds, the energy
of the created particles can be related to dissipation in vacuum in a
similar way.

The probability of particle creation 
(\ref{stochastic probability}) is
a fluctuating quantity, due to the functional dependence
on the stochastic perturbation $h_{ab}$. 
We may compute its averaged
value $\langle P[h;g] \rangle_c$, which [neglecting the higher order
corrections in Eq.~(\ref{stochastic probability})] is given by
\be
\langle P[h;g] \rangle_c = P[\langle h \rangle_c;g]+
\int\! d^4x\, d^4y \, \sqrt{- g(x)}\sqrt{- g(y)}\: 
N^{abcd}[g](x,y)\,
\langle h^{\rm f}_{ab}(x) h^{\rm f}_{cd}(y) \rangle_c,
\label{mean P}
\ee
where $h_{ab}^{\rm f} \equiv h_{ab} -\langle h_{ab} \rangle_c$. 
The first term in the right hand side of Eq.~(\ref{mean P}) 
is the probability of particle creation 
(or one half of the number of created particles) that one would
obtain in the spacetime
$({\cal M},g_{ab} \!+\! \langle h_{ab} \rangle_c)$.
The second term will be greater than zero when stress-energy
fluctuations are present\footnote{Except in some rare cases, 
for which 
$N^{abcd}(x,y)$ is not strictly positive
definite and $\langle h^{\rm f}_{ab}(x) h^{\rm f}_{cd}(y) \rangle_c$ 
is such that it ``hits'' the zero eigenvalue.}
since, from the Einstein-Langevin equation, this implies 
$\langle h^{\rm f}_{ab}(x) h^{\rm f}_{cd}(y) \rangle_c \!\neq \! 0$.
Note that, when this is the case, from the fluctuation-dissipation 
relation of subsection \ref{subsub:zero T}, the vacuum 
dissipation kernel will be also non-vanishing.
Hence, metric fluctuations induced by matter stress-energy
fluctuations generally increase the mean value of the number of
created particles with respect to the same quantity 
in the ``perturbed'' semiclassical spacetime
$({\cal M},g_{ab} \!+\! \langle h_{ab} \rangle_c)$.

The above result for the total probability of particle creation 
and number of created particles can
be easily generalized to the case of a massless conformally coupled
scalar field and 
a conformally stationary semiclassical background. 
When this background is a spatially
flat RW universe \cite{cv96}, 
performing conformal transformations in the metric
perturbations and in the kernels as in 
Eq.~(\ref{conformal relations for kernels}), 
one gets expressions analogous to (\ref{prob in Minkowski}), 
(\ref{relation in Minkowski}) and 
(\ref{energy of particles in Minkowski}) with 
$N_2 \!=\! D_2 \!=\! 0$ (see Refs.~\cite{cv94,cv96,mv98} 
for more details).

%%%%%%%%%%%%%%%%%%%%%%%%%%%%%%%%%%%%%%%%

\section{Conclusions}
\label{sec:conclu}

%%%%%%%%%%%%%%%%%%%%%%%%%%%%%%%%%%%%%%%%

In the first part of this paper, 
we have shown how a consistent stochastic semiclassical
theory of gravity can be formulated. This theory is a perturbative
generalization of semiclassical gravity which describes the back
reaction of the lowest order stress-energy fluctuations of quantum
matter fields on the gravitational field through the semiclassical
Einstein-Langevin equation. We have shown that this equation 
can be formally derived with a method
based on the influence functional of Feynman and Vernon, where one
considers the metric field as the ``system'' of interest and the
matter fields as part of its ``environment'' \cite{hu89}. 
Our approach clarifies the physical meaning of the semiclassical 
Langevin-type equations previously derived with the same functional
method 
\cite{calzettahu,humatacz,husinha,cv96,lomb-mazz,cv97,ccv97,%
campos-hu,campos-hu2,calver98}, since it links the source of stochastic
fluctuations to quantum matter stress-energy fluctuations, and allows
to formulate the theory in a general way.
At the same time, we have also developed a method to compute the
semiclassical Einstein-Langevin equation using dimensional
regularization. This provides an alternative and more direct way of
computing the equation with respect to the previous calculations,
based on a specific evaluation of the effective action of Feynman and
Vernon \cite{calzettahu,humatacz,husinha,cv96,lomb-mazz,cv97,ccv97,%
campos-hu,campos-hu2,calver98}.
In a subsequent paper \cite{paperII}, we shall apply this method to
solve the Einstein-Langevin equation around some simple solutions of
semiclassical gravity.

The second part of the paper was devoted to the existence
of fluctuation-dissipation relations and to particle creation in
the context of stochastic semiclassical gravity.
When the background solution of semiclassical gravity consists of a
stationary spacetime and a scalar field in a thermal equilibrium state,
we have identified a dissipation kernel in the Einstein-Langevin
equation which is related to the noise kernel by a
fluctuation-dissipation relation. The same relation was
previously derived by Mottola \cite{mottola} using a linear response
theory approach. We have also generalized this result to the case of a
conformal scalar field in a conformally stationary background solution
of semiclassical gravity.

Our analysis seems to indicate that for a fluctuation-dissipation
relation to be present in stochastic semiclassical gravity, the
semiclassical background solution must satisfy certain conditions. 
In this paper we have just analyzed the simplest cases for which such
a relation exists. Further work must be done to investigate whether a
similar relation is present in other situations of physical
interest, 
such as black hole backgrounds \cite{hu99,hu-raval-sinha,campos-hu2},
or non-conformal fields in RW backgrounds in the instantaneous 
vacua or the thermal states defined in Ref.~\cite{weiss86}.

We have also studied particle creation by stochastic metric
perturbations in stationary and conformally stationary (for conformal
matter fields in this latter case) background solutions of
semiclassical gravity. We have expressed the
total probability of particle creation and the number of created
particles (the expectation value of the number operator for 
``out'' particles in the ``in'' vacuum) in terms of the vacuum noise 
kernel. We have shown that the averaged value of those quantities
is enhanced by the presence of stochastic metric fluctuations.
In the particular cases of a Minkowski
background and a conformal field in a spatially flat RW background, 
the energy of the created particles can 
be expressed in terms of the vacuum dissipation kernels.

It should be stressed that the concept of particle creation is only
well defined when the solutions of the Einstein-Langevin equation
vanish in the ``remote past''
and in the ``far future'' (at least, ``asymptotically''). 
However, 
there can be physically meaningful solutions of the Einstein-Langevin
equation that do not satisfy these rather strong conditions.
In this case, 
vacuum noise and dissipation in stochastic semiclassical gravity can
include effects that are not associated to particle creation.

%%%%%%%%%%%%%%%%%%%%%%%%%%%%%%%%%%%
%                                 %
%        Acknowledgments          %
%                                 %
%%%%%%%%%%%%%%%%%%%%%%%%%%%%%%%%%%%

\acknowledgments

We are grateful to Esteban Calzetta, Jaume Garriga,
Bei-Lok Hu, Ted Jacobson and Albert Roura
for very helpful suggestions and discussions. 
This work has been partially supported by the 
CICYT Research Project number
\mbox{AEN95-0590}, and the European Project number
\mbox{CI1-CT94-0004}.

%%%%%%%%%%%%%%%%%%%%%%%%%%%%%%%%%%%%%%%%

\appendix

%%%%%%%%%%%%%%%%%%%%%%%%%%%%%%%%%%%%%%%%

%%%%%%%%%%%%%%%%%%%%%%%%%%%%%%%%%%%%%%%%

\section{Expansion of the stress-energy tensor
around a background metric}
\label{sec:expansions of tensors}

%%%%%%%%%%%%%%%%%%%%%%%%%%%%%%%%%%%%%%%%

The expansion of the stress-energy tensor functional
\[
T^{ab}[g,\Phi_{n}]\equiv \bigtriangledown^a \Phi_{n} \!
\bigtriangledown^b \!\Phi_{n}- {1\over 2}\, g^{ab}\hspace{-0.1ex} 
\bigtriangledown^{c}\!\Phi_{n}\! \bigtriangledown_{\!c}\!\Phi_{n} 
-{1\over 2}\, g^{ab}\hspace{0.2ex} m^2 \Phi_{n}^2 
+\xi \left( g^{ab} \Box
-\bigtriangledown^a \!\bigtriangledown^b
+G^{ab} \right)  \Phi_{n}^2,
\]
around a background metric $g_{ab}$ is given by
\bea
&&T^{ab}[g+h,\Phi_{n}]=T^{ab}[g,\Phi_{n}]+
T^{{\scriptscriptstyle (1)}\hspace{0.1ex} ab}[g,\Phi_{n};h]
+0(h^2), \hspace{3ex} \mbox{with}  \nn \\
&&T^{{\scriptscriptstyle (1)}\hspace{0.1ex} ab}[g,\Phi_{n};h]=
-T^{ac}[g,\Phi_{n}]\, h^b_c-T^{bc}[g,\Phi_{n}]\, h^a_c
-{1 \over 2}\, \bigl( 
\bigtriangledown^{c}\Phi_{n}\! \bigtriangledown_{\!c}\!\Phi_{n}
+m^2 \Phi_{n}^2 \hspace{0.1ex}\bigr) \, h^{ab}
+ {1 \over 2}\, g^{ab} \bigtriangledown^c\! \Phi_{n}\!
\bigtriangledown^d\! \Phi_{n} \, h_{cd} 
 \nn \\
&& \hspace{18ex}
+\, {\xi \over 2} \left[ -R\, h^{ab}+g^{ab} R^{cd}h_{cd}
+\bigtriangledown^c \!\bigtriangledown^a \! h^b_c
+\bigtriangledown^c \!\bigtriangledown^b \! h^a_c
-\bigtriangledown^a \!\bigtriangledown^b \! h^c_c
-\Box h^{ab} 
+ g^{ab} \,\bigl( \Box h^c_c \hspace{-0.1ex}-\hspace{-0.1ex}
\bigtriangledown^c \!\bigtriangledown^d \! h_{cd}
\hspace{0.2ex} \bigr) 
 \right. \nn \\
&& \hspace{18ex} \left.
+\, \bigl( \bigtriangledown^a h^b_c
\hspace{-0.1ex}+\hspace{-0.1ex} \bigtriangledown^b h^a_c
\hspace{-0.1ex}-\hspace{-0.1ex}
\bigtriangledown_{\!c}\hspace{0.2ex} h^{ab}\hspace{-0.1ex}
-2\hspace{0.1ex} g^{ab}\hspace{-0.2ex} 
\bigtriangledown^d \! h_{cd}
+g^{ab}\hspace{-0.2ex} \bigtriangledown_{\!c}\! h^d_d
\hspace{0.2ex} \bigr)  \bigtriangledown^c
+2\hspace{0.1ex} h^{ab}\, \Box
- 2\hspace{0.1ex} g^{ab}\hspace{0.2ex} 
h_{cd} \bigtriangledown^c\! \bigtriangledown^d
\hspace{0.2ex}
\right] \Phi_{n}^2,  \nn
\eea
where the covariant derivatives and curvature tensors 
are those of the metric $g_{ab}$, and indices are raised with
inverse background metric $g^{ab}$.

%%%%%%%%%%%%%%%%%%%%%%%%%%%%%%%%%%%%%%%%

\section{Hamiltonian operator in a stationary spacetime
and thermal Wick's theorem}
\label{sec:Hamiltonian}

%%%%%%%%%%%%%%%%%%%%%%%%%%%%%%%%%%%%%%%%

In this appendix, we construct the Hamiltonian or energy
operator for a quantum scalar field in a stationary spacetime.
For a more rigorous mathematical treatment, see Ref.~\cite{kay} and,
for the particular case of a static spacetime, see Ref.~\cite{fulling}.
We also show how this construction can be generalized for a conformal
scalar field in a conformally stationary spacetime. 
Using this Hamiltonian to define a thermal density operator, we shall
see how thermal four-point functions can be expressed in terms of
thermal two-point functions (``thermal Wick's theorem'').

Let $({\cal M},g_{ab})$ be a $n$ dimensional stationary spacetime, 
that is, a
spacetime with a global timelike Killing vector field
$\zeta^a \equiv (\partial / \partial t)^a$, and consider a linear real
scalar field $\Phi_{n}$ on it. Assuming that the spacetime is globally
hyperbolic, we can foliate it by a family of Cauchy hypersurfaces
$\Sigma_{t}$, labeled by the Killing time $t$ (hypersurfaces
of constant $t$), and give coordinates to each
point of the spacetime $x^{\mu}\!=\!(t,{\bf x})$, 
where ${\bf x}\equiv (x^i)$ are local
coordinates on each of these hypersurfaces. Let $n^a$ be the 
future directed unit
({\it i.e.}, $n_a n^a=-1$ and $n^t > 0$)
vector field normal to each hypersurface $\Sigma_{t}$. 
The induced metric on each $\Sigma_{t}$ by the spacetime metric
is  $q_{ab} \equiv g_{ab}+n_{a} n_{b}$ \cite{wald84}, then 
$q^a_b$ is a projector orthogonal to $n^a$. We can decompose
the Killing vector into its normal and tangential parts to each
$\Sigma_{t}$: $\zeta^a \!=\! N n^a\!+\!N^a$, where 
$N \equiv -\zeta^a n_{a}$
and $N^a \equiv q^a_b \, \zeta^b$ 
are, respectively, the lapse function and the shift vector.
In the basis associated to 
the coordinate system $\{ x^{\mu} \}$, the components
$g_{\mu \nu}$ of the metric are independent of $t$ and
can be written as
$g_{tt}=-N^2+N_i N^i$, $g_{ti}=N_i$,
$g_{ij}=q_{ij}$, with $N_i=q_{ij} N^j$. 
One can also write 
$\sqrt{-g}= N \sqrt{q}$, where 
$q \equiv \det \hspace{0.2ex} (q_{ij})$.

To construct the classical Hamiltonian,
we write the Lagrangian density as
\be
{\cal L}_n={1\over2} \, \sqrt{q}\, N  \left[ 
(n^{\mu} \partial_{\mu}\Phi_{n})^2
-q^{ij}\partial_i \Phi_{n} \partial_j \Phi_{n} 
-(m^2+ \xi R )\hspace{0.2ex} \Phi_{n}^2 \right],
\ee
where $n^t\!=\!1/N$,
$n^i\!=\!-N^i/N$ and $q^{ij}$ is the inverse of $q_{ij}$,
$q_{ik}\hspace{0.3ex} q^{kj}\!= \delta_i^j$.
The momentum canonical conjugate to $\Phi_{n}$ is
$\Pi_{n} = 
\sqrt{q}\: n^{\mu} \partial_{\mu}\Phi_{n}$ 
and the Hamiltonian density is constructed as usual,
${\cal H}_n= \Pi_{n}\, \partial_t \Phi_{n}-{\cal L}_n$, 
from which the Hamiltonian functional on the 
hypersurface $\Sigma_{t}$ is given by
$H(t)\!=\! \int_{\Sigma_{t}}  d^{n-1}{\bf x} \: {\cal H}_n(x)$.
Integrating by parts and dropping surface terms, 
we get 
\bea
&&H(t)={1\over2} \hspace{-0.2ex} 
\int_{\Sigma_{t}} \hspace{-0.3ex}
d^{n-1}{\bf x} \, \sqrt{q} \,
\Biggl\{ -\Phi_{n}  \hspace{-0.2ex}
\left[ (D^i N) D_i+N\, ( \Delta -m^2- \xi R )
\right] \hspace{-0.2ex} \Phi_{n}
-\Phi_{n} \hspace{0.2ex} (D_i N^i+N^i D_i) \hspace{-0.4ex}
\left( {\Pi_{n} \over \sqrt{q}} \right)   \nn \\
&&\hspace{26ex} 
+\, {\Pi_{n} \over \sqrt{q}}\, N^i D_i \Phi_{n}
+N \hspace{-0.2ex}
\left( {\Pi_{n} \over \sqrt{q}} \right)^{\!\!2} 
\hspace{0.2ex} \Biggr\},
\label{Hamiltonian}
\eea
where $D_i$ is the covariant derivative on the $n\!-\!1$ dimensional
Riemannian spaces $(\Sigma_{t},q_{ij})$ 
(associated to the metric $q_{ij}$), and 
$\Delta \equiv D^i D_i$ is the associated Laplace-Beltrami operator. 
For a field $\Phi_{n}$ and its momentum
conjugate $\Pi_{n}$ satisfying the Hamiltonian equations of motion,
this Hamiltonian is a conserved quantity, {\it i.e.},
it is independent of $t$.

The analogous quantity in
the Lagrangian formalism is the canonical energy functional,
which is defined in terms of the canonical
stress-energy tensor functional,
\be
T_{ab}^{\rm can}[g,\Phi_{n}] \equiv 
\bigtriangledown_{\!a}\Phi_n \!
\bigtriangledown_{\!b}\!\Phi_n
- {1\over 2}\, g_{ab} 
\bigtriangledown^{c}\!\Phi_n \! \bigtriangledown_{\!c}\!\Phi_n 
-{1\over 2}\, g_{ab}\, (m^2+ \xi R)\, \Phi_n^2 , 
\label{canonical s-e}
\ee
as
$E_{\rm can} \!\equiv \! \int_{\Sigma} 
d \Sigma  \; n^a \zeta^b \, T_{ab}^{\rm can}[g,\Phi_{n}]$,
where $\Sigma$ is a Cauchy hypersurface, $n^a$ is the future directed 
unit vector field normal to $\Sigma$, 
and $d \Sigma$ is the invariant volume element on $\Sigma$
constructed with the metric induced by $g_{ab}$.
By Noether's theorem \cite{wald84}, this 
energy functional is conserved, {\it i.e.}, it is 
independent of the choice of 
Cauchy hypersurface $\Sigma$, when $\Phi_{n}$ satisfies the
Klein-Gordon equation.
Choosing $\Sigma_t$ as the Cauchy hypersurface, 
we can obtain an expression 
for $E_{\rm can}$ after the substitution of 
$\Pi_{n}$ by $\sqrt{q}\: n^{\mu} \partial_{\mu}\Phi_{n}$ 
in the Hamiltonian (\ref{Hamiltonian}).
Note that we can also introduce an energy functional 
$E \equiv \int_{\Sigma} 
d \Sigma  \; n^a \zeta^b \, T_{ab}[g,\Phi_{n}]$,
where $T_{ab}$ is the stress-energy tensor functional 
(\ref{class s-t}) \cite{wald84}. 
For a field $\Phi_{n}$ satisfying the Klein-Gordon
equation, this is also a conserved quantity. 
However, choosing a Cauchy hypersurface $\Sigma_t$, 
one can show that
$n^a \zeta^b \, (T_{ab}-T^{\rm can}_{ab})$ is a divergence on
the space $(\Sigma_t, q_{ij})$ and, thus, dropping surface terms, 
we have $E=E_{\rm can}$.

We can now formally construct the Hamiltonian ``operator'' in
the Heisenberg picture simply by replacing $\Phi_{n}$ and
$\Pi_{n}$ by their corresponding operators 
$\hat{\Phi}_{n}$ and $\hat{\Pi}_{n}$ in (\ref{Hamiltonian}) and using,
as always, a Weyl ordering prescription for the operators.
This operator is a conserved quantity, that is, it is
independent of the time $t$; 
therefore, it is equal to the Hamiltonian operator in the
Schr\"{o}dinger picture and we simply denote it by $\hat{H}$.
Since the momentum operator in the Heisenberg picture satisfies 
$\hat{\Pi}_{n}\!=\!\sqrt{q}\, n^{\mu} \partial_{\mu}\hat{\Phi}_{n}$, 
this Hamiltonian operator can also be obtained from the 
canonical energy functional 
[hence, $\hat{H}$ represents also a conserved energy operator].
Taking into account that the field
operator $\hat{\Phi}_{n}$ satisfies the Klein-Gordon equation, 
we find
\be
\hat{H}= {1\over4} \hspace{-0.2ex} 
\int_{\Sigma_{t}} \hspace{-0.3ex}
d^{n-1}{\bf x} \, \sqrt{q} \, {1 \over N}
\left[ \Bigl\{ \partial_t \hat{\Phi}_{n}
\hspace{0.3ex} , \hspace{0.3ex}
(\partial_t \hat{\Phi}_{n}- N^i \partial_i \hat{\Phi}_{n})
\Bigr\} - \Bigl\{ \hat{\Phi}_{n} 
\hspace{0.3ex} , \hspace{0.3ex}
\partial_t \hspace{0.3ex}
(\partial_t \hat{\Phi}_{n}- N^i \partial_i \hat{\Phi}_{n})
\Bigr\} \right].
\label{Hamiltonian operator}
\ee

In this case, there exists a natural Fock representation based on a
decomposition of the field operator $\hat{\Phi}_{n}$ in terms of a
complete set of modes 
$\{ u_{k_{\mbox{}_{\scriptstyle n}}}\hspace{-0.4ex}(x) \}$, 
solution of the Klein-Gordon
equation, which have 
positive frequency with respect to the Killing vector
$\zeta^a \equiv (\partial / \partial t)^a$:
$\partial_t u_{k_{\mbox{}_{\scriptstyle n}}}\hspace{-0.4ex}(x)
=-i \hspace{0.3ex} \omega_{k} \hspace{0.3ex} 
u_{k_{\mbox{}_{\scriptstyle n}}}\hspace{-0.4ex}(x)$,
with $\omega_{k}\!>\!0$. The label $k$ of each mode must in general be
understood as representing a set of discrete or continuous indices,
and, thus, the summations over $k$ represent either a discrete
sum or an integral with some suitable measure (or a combination of
these two possibilities). We assume that these modes have the
same physical dimensions as the field $\Phi_{n}$ (this is the reason
why we put a subindex $n$).
These modes have to be orthonormal with respect to the inner product
$(\phi_1 , \phi_2 ) \equiv -i \int_{\Sigma} 
d\Sigma \: n^a \hspace{-0.3ex}
\left( \phi_1 \partial_a \phi_2^{\displaystyle\,\ast}
-\phi_2^{\displaystyle\,\ast} \partial_a \phi_1
\right)$,
which is independent of the Cauchy hypersurface $\Sigma$ 
when $\phi_1$ and $\phi_2$ are solutions
of the Klein-Gordon equation \cite{birrell,dewitt},
{\it i.e.}, 
$(u_{k_{\mbox{}_{\scriptstyle n}}},
u_{l_{\mbox{}_{\scriptstyle n}}} ) =\delta_{kl}$ and 
$(u_{k_{\mbox{}_{\scriptstyle n}}},
u_{l_{\mbox{}_{\scriptstyle n}}}^{\displaystyle \ast} ) =0$.

The field operator can then be written as
\be
\hat{\Phi}_{n}(x)=\sum_{k} \left[ 
u_{k_{\mbox{}_{\scriptstyle n}}}\hspace{-0.4ex}(x)\, \hat{a}_k
+u^{\displaystyle \ast}
  _{k_{\mbox{}_{\scriptstyle n}}}\hspace{-0.4ex}(x)\,
\hat{a}_k^{\dag}
\right],
\label{field operator}
\ee
where $\hat{a}_k^{\dag}$ and $\hat{a}_k$ are creation and annihilation
operators on the Fock space associated to this mode decomposition, 
which satisfy the usual commutation relations \cite{birrell,dewitt}.
Using these commutation relations and the orthonormality conditions
for the modes evaluated on $\Sigma_{t}$, substituting 
(\ref{field operator}) into (\ref{Hamiltonian operator}), 
one finds the Fock space
representation of the formal Hamiltonian ``operator''
$\hat{H}\!=\!\sum_{k} \omega_{k} \hspace{0.2ex}
(\hat{a}_k^{\dag}\hat{a}_k +{\textstyle \frac{1}{2}})$.
We can make this last expression well defined by subtraction of
the ``divergent'' constant c-number $\sum_{k} (\omega_{k}/2)$,
that is, we can introduce a renormalized 
Hamiltonian operator as
\be
\hat{H}_{\!R}=\sum_{k} \omega_{k} \,\hat{a}_k^{\dag} \hat{a}_k.
\label{Hamiltonian operator on Fock space 2}
\ee
Note that $\hat{H}_{\!R}$ is given by an expression similar to 
(\ref{Hamiltonian operator}), but adding a normal ordering
prescription for the operators $\hat{a}_k$ and $\hat{a}_k^{\dag}$,
or, equivalently (dropping surface terms), 
by 
$\hat{H}_{\!R}= 
\int_{\Sigma} 
d \Sigma  \: n^a \zeta^b \hspace{-0.7ex} : \hspace{-0.5ex}
\hat{T}_{n_{\scriptstyle ab}}[g] \hspace{-0.6ex}: \hspace{0.4ex}
= \int_{\Sigma} 
d \Sigma  \: n^a \zeta^b \hspace{-0.7ex} : \hspace{-0.5ex}
\hat{T}_{n_{\scriptstyle ab}}^{\rm can}[g] 
\hspace{-0.6ex} : \hspace{0.4ex}$,
where $\hat{T}_{n_{\scriptstyle ab}}[g]$ is defined in
(\ref{regul s-t}), $\hat{T}_{n_{\scriptstyle ab}}^{\rm can}[g]$
is analogously defined after (\ref{canonical s-e}),
and $: \; :$ means normal ordering \cite{dewitt,mostepanenko}.
The vacuum and the many-particle states of the Fock space are
eigenstates of this Hamiltonian operator with zero and
positive eigenvalues, respectively, 
(given by the sum of the $\omega_{k}$'s
corresponding to the particle contents of the state).

From (\ref{field operator}) and 
$\hat{\Pi}_{n}\!=\!\sqrt{q}\, n^{\mu} \partial_{\mu}\hat{\Phi}_{n}$,
using the positive frequency condition and
\be
[\hspace{0.1ex} \hat{H}_{\!R}
\hspace{0.2ex},\hat{a}_k \hspace{0.1ex}]= - \omega_k \, \hat{a}_k,
\hspace{10ex}
[\hspace{0.1ex} \hat{H}_{\!R}
\hspace{0.2ex},\hat{a}_k^{\dag} \hspace{0.1ex}]= 
\omega_k \, \hat{a}_k^{\dag},
\label{comm rels}
\ee
we get
$\partial_{t}\hat{\Phi}_{n}\! =\! i \, [\hspace{0.1ex} \hat{H}_{\!R}
\hspace{0.2ex}, \hat{\Phi}_{n}\hspace{0.1ex}]$
and
$\partial_{t}\hat{\Pi}_{n}\! =\! i \, [\hspace{0.1ex} \hat{H}_{\!R}
\hspace{0.2ex}, \hat{\Pi}_{n}\hspace{0.1ex}]$.
These are the Heisenberg equations of motion, which are equivalent to
the Klein-Gordon equation for the operator $\hat{\Phi}_{n}$. 
From these equations, we see that 
the operator 
\mbox{
$\exp \hspace{0.2ex}\bigl(-i \hat{H}_{\!R}(t\!-\!t^\prime)\bigr)$} 
generates the time evolution of operators in the Heisenberg picture.

\subsection{Conformal field in a conformally stationary spacetime}

Let us now consider a massless conformally coupled real scalar field 
$\bar{\Phi}_{n}$ in a $n$ dimensional spacetime 
$({\cal M},\overline{g}_{ab})$,  assumed to be conformally
stationary and globally hyperbolic. The action 
$S_m[\bar{g},\bar{\Phi}_{n}]$ for the field is given by 
(\ref{scalar action}) with $m\!=\!0$ and 
$\xi\!=\!\xi(n)\!\equiv \!(n-2)/[4(n-1)]$. 
In this case, the spacetime has a global timelike conformal 
Killing vector field
$\zeta^a \equiv (\partial / \partial t)^a$, which satisfies
$\bbderiv_{\!a}\!\zeta_{b}+\bbderiv_{\!b}\!
\zeta_{a}=(2/n)\, \bbderiv^{c}\hspace{-0.5ex}\zeta_{c}
\:\overline{g}_{ab}$, where $\deriv_{\!a}$ 
is the covariant derivative associated to the metric 
$\overline{g}_{ab}$. The metric $\overline{g}_{ab}$ is conformally related
to a stationary metric $g_{ab}$: 
$\overline{g}_{ab}(x)=e^{2 \varpi(x)} g_{ab}(x)$.
The foliation of the spacetime by Cauchy
hypersurfaces $\Sigma_{t}$ and the coordinates  
$x^{\mu}\!=\!(t,{\bf x})$ are introduced as above.

Given a Cauchy hypersurface $\Sigma$, with unit normal $\bar{n}^a$
(as above, we take $\bar{n}^t \! > \! 0$), we can introduce the energy
functional as
$E \equiv \int_{\Sigma} \;
\overline{\!\!d \Sigma \!}  \;\, \bar{n}^a \zeta^b \, 
T_{ab}[\bar{g},\bar{\Phi}_{n}]$, 
where $\,\,\overline{\!\!d \Sigma \!}\,$ is the invariant
volume element constructed with the metric on $\Sigma$ induced by the
metric $\overline{g}_{ab}$. Given that the stress-energy tensor 
$T_{ab}[\bar{g},\bar{\Phi}_{n}]$ is traceless when 
the field $\bar{\Phi}_{n}$ satisfies the Klein-Gordon equation, 
it is easy to see from the equation for the conformal Killing vector
$\zeta^a$ 
that this energy functional is conserved.
In fact, choosing a hypersurface $\Sigma_t$ to evaluate this energy,
and introducing 
$\Phi_n \equiv e^{(n-2)\varpi /2} \hspace{0.2ex} \bar{\Phi}_{n}$,
it is easy to see \cite{ford75} that
\be
E = \int_{\Sigma_{t}} \hspace{-0.4ex}
d^{n-1}{\bf x} \, \sqrt{q} \:  
n^a \zeta^b \, T_{ab}[g,\Phi_{n}],
\label{energy 3}
\ee
where $n^a$ and $q_{ij}$ are constructed with the metric $g_{ab}$.
Thus, $E$ is equal to the energy functional 
for the field $\Phi_{n}$
in the stationary spacetime $({\cal M},g_{ab})$.

Using the ``natural'' Fock representation, based on the decomposition
of the field operator $\hat{\Phi}_{n}[\bar{g}]$ in terms of modes
$\bar{u}_{k_{\mbox{}_{\scriptstyle n}}}\hspace{-0.4ex}(x)=
e^{-(n-2)\varpi(x)/2} \hspace{0.2ex}
u_{k_{\mbox{}_{\scriptstyle n}}}\hspace{-0.4ex}(x)$,
we can construct the
renormalized energy operator in the Heisenberg picture,
$\hat{E}_{\!R}[\bar{g}]$,
associated to the above energy functional $E$.
Here, as above, 
$\{ u_{k_{\mbox{}_{\scriptstyle n}}}\hspace{-0.4ex}(x) \}$
is a complete set of modes, solution 
of the Klein-Gordon equation in the stationary spacetime 
$({\cal M},g_{ab})$, which are of positive frequency with respect to
$\zeta^a \equiv (\partial / \partial t)^a$.
Dropping surface terms, we get 
$\hat{E}_{\!R}[\bar{g}]=
\sum_{k} \omega_{k} \,\hat{a}_k^{\dag} \hat{a}_k$, where
$\hat{a}_k^{\dag}$ and $\hat{a}_k$ are creation and annihilation
operators on the Fock space associated to these conformal modes.

Alternatively, one can perform the transformation 
$\bar{\Phi}_{n} \equiv e^{-(n-2)\varpi /2} \hspace{0.2ex} \Phi_n$
in the scalar field action, which is then transformed to 
$S_m[g,\Phi_n]$, 
and construct the Hamiltonian associated to this transformed action,
which is given by the above expressions for the stationary case.
This is equivalent to making a canonical transformation in the
Hamiltonian formulation of the theory. 
One then introduces an operator 
$\hat{\Phi}_{n}[g] \equiv        
e^{(n-2)\varpi/2} \hspace{0.2ex} \hat{\Phi}_{n}[\bar{g}]$,
which can be identified as the field
operator in the Heisenberg picture 
in the stationary spacetime 
$( {\cal M}, g_{ab})$.
The associated Hamiltonian operator 
can be identified 
with the operator $\hat{H}_{\!R}[g]$ constructed above
(and, obviously, it coincides with $\hat{E}_{\!R}[\bar{g}]$).
Note that this Hamiltonian or energy operator generates the time
evolution of the operator $\hat{\Phi}_{n}[g]$ rather than that of 
the ``physical'' field $\hat{\Phi}_{n}[\bar{g}]$.
A generalization of this last approach has been used in 
Ref.~\cite{weiss86} for scalar fields 
with arbitrary mass and arbitrary coupling to the curvature
in a RW spacetime
to construct a time-dependent 
Hamiltonian operator whose ground state at each fixed
instant of time is a Hadamard state. 
A similar construction starting with the above energy functional
is given in Ref.~\cite{mostepanenko}.
In the massless conformally
coupled case, these time-dependent Hamiltonian constructions 
reduce to the construction sketched in this appendix.

\subsection{Wick's theorem for thermal states}

From the Hamiltonian operator 
(\ref{Hamiltonian operator on Fock space 2}) (here, we drop 
the subindex $R$), we can define a state of thermal equilibrium for
the scalar field as in (\ref{thermal state}).
Following partially the proof presented in the appendix of 
Ref.~\cite{lebellac}, 
we shall next show how Wick's theorem can be generalized for the
associated thermal $N$-point functions. 
First, note that, from (\ref{comm rels}), 
\be
\hat{a}_k^{(\a )} e^{-\beta \hat{H}} =
e^{-\a \, \beta \omega_k} e^{-\beta \hat{H}} \, \hat{a}_k^{(\a )},
\label{thermal prop}
\ee
where $\a \!=\! +,-$, $\hat{a}_k^{(+)} \!\equiv \! \hat{a}_k$ and 
$\hat{a}_k^{(-)} \!\equiv \! \hat{a}_k^{\dag}$.
Using this and the cyclic property of the trace, we get
\be
\left \langle \hat{a}_k^{(\a )} \hat{a}_l^{(\g )}
\right \rangle_{ \mbox{}_{\! \scriptscriptstyle T}} =
{1 \over 1-e^{-\a \, \beta \omega_k} } 
\left[ \hspace{0.1ex} \hat{a}_k^{(\a )} ,
\hat{a}_l^{(\g )} \hspace{0.1ex} \right],
\label{thermal exp val}
\ee
where we have used the commutator
$[ \hspace{0.1ex} \hat{a}_k^{(\a )} ,
\hat{a}_l^{(\g )} \hspace{0.1ex} ]$
to represent either 
$\delta_{kl}$, $-\delta_{kl}$ or $0$ (such commutator does not
represent an operator in the last equation). 
Writing the field operator $\hat{\Phi}_{n}(x)$ in terms of the 
operators $\hat{a}_k^{(\a )}$, the associated four-point 
thermal functions can be expressed in terms of 
$\langle \hat{a}_k^{(\a )} \hat{a}_l^{(\g )}
\hat{a}_r^{(\d )} \hat{a}_s^{(\s )} 
 \rangle_{\mbox{}_{\scriptscriptstyle T}}$.
Taking into account that 
the commutator $[ \hspace{0.1ex} \hat{a}_k^{(\a )} ,
\hat{a}_l^{(\g )} \hspace{0.1ex}]$ is a c-number, 
one has the following identity
\be
\hat{a}_k^{(\a )} \hat{a}_l^{(\g )}
\hat{a}_r^{(\d )} \hat{a}_s^{(\s )} =
\left[ \hspace{0.1ex} \hat{a}_k^{(\a )} \hspace{-0.1ex},
\hat{a}_l^{(\g )} \hspace{0.1ex} \right] 
\hat{a}_r^{(\d )} \hat{a}_s^{(\s )}
\!+\! \left[ \hspace{0.1ex} \hat{a}_k^{(\a )} \hspace{-0.1ex},
\hat{a}_r^{(\d )} \hspace{0.1ex} \right] 
\hat{a}_l^{(\g )} \hat{a}_s^{(\s )}
\!+\!\left[ \hspace{0.1ex} \hat{a}_k^{(\a )} \hspace{-0.1ex},
\hat{a}_s^{(\s )} \hspace{0.1ex} \right] 
\hat{a}_l^{(\g )} \hat{a}_r^{(\d )}
\!+ \hat{a}_l^{(\g )}
\hat{a}_r^{(\d )} \hat{a}_s^{(\s )} \hat{a}_k^{(\a )}.
\label{ident}
\ee
On the other hand, from (\ref{thermal prop})
and the cyclic property of the trace, we have
\be
\left \langle \hat{a}_l^{(\g )}
\hat{a}_r^{(\d )} \hat{a}_s^{(\s )} \hat{a}_k^{(\a )} 
\right \rangle_{ \mbox{}_{\! \scriptscriptstyle T}} =
e^{-\a \, \beta \omega_k}  
\left \langle \hat{a}_k^{(\a )} \hat{a}_l^{(\g )}
\hat{a}_r^{(\d )} \hat{a}_s^{(\s )} 
\right \rangle_{ \mbox{}_{\! \scriptscriptstyle T}}.
\label{ident 2}
\ee
Using the last two equations, we get
\be
\left \langle \hat{a}_k^{(\a )} \hat{a}_l^{(\g )}
\hat{a}_r^{(\d )} \hat{a}_s^{(\s )} 
\right \rangle_{ \mbox{}_{\! \scriptscriptstyle T}}=
{1 \over ( 1-e^{-\a \, \beta \omega_k} ) } \,
\left\{ 
\left[ \hspace{0.1ex} \hat{a}_k^{(\a )} \hspace{-0.1ex},
\hat{a}_l^{(\g )} \hspace{0.1ex} \right]
\left \langle \hat{a}_r^{(\d )} \hat{a}_s^{(\s )}
\right \rangle_{ \mbox{}_{\! \scriptscriptstyle T}} 
\!+\! \left[ \hspace{0.1ex} \hat{a}_k^{(\a )} \hspace{-0.1ex},
\hat{a}_r^{(\d )} \hspace{0.1ex} \right]
\left \langle \hat{a}_l^{(\g )} \hat{a}_s^{(\s )}
\right \rangle_{ \mbox{}_{\! \scriptscriptstyle T}} 
+ \left[ \hspace{0.1ex} \hat{a}_k^{(\a )} \hspace{-0.1ex},
\hat{a}_s^{(\s )} \hspace{0.1ex} \right]
\left \langle \hat{a}_l^{(\g )} \hat{a}_r^{(\d )}
\right \rangle_{ \mbox{}_{\! \scriptscriptstyle T}} 
\right\} \!,
\ee
which, from (\ref{thermal exp val}), yields
\be
\left \langle \hat{a}_k^{(\a )} \hat{a}_l^{(\g )}
\hat{a}_r^{(\d )} \hat{a}_s^{(\s )} 
\right \rangle_{ \mbox{}_{\! \scriptscriptstyle T}}=
\left \langle \hat{a}_k^{(\a )} \hat{a}_l^{(\g )}
\right \rangle_{ \mbox{}_{\! \scriptscriptstyle T}}
\left \langle \hat{a}_r^{(\d )} \hat{a}_s^{(\s )}
\right \rangle_{ \mbox{}_{\! \scriptscriptstyle T}}
+\left \langle \hat{a}_k^{(\a )} \hat{a}_r^{(\d )}
\right \rangle_{ \mbox{}_{\! \scriptscriptstyle T}}
\left \langle \hat{a}_l^{(\g )} \hat{a}_s^{(\s )}
\right \rangle_{ \mbox{}_{\! \scriptscriptstyle T}}
+\left \langle \hat{a}_k^{(\a )} \hat{a}_s^{(\s )}
\right \rangle_{ \mbox{}_{\! \scriptscriptstyle T}}
\left \langle \hat{a}_l^{(\g )} \hat{a}_r^{(\d )}
\right \rangle_{ \mbox{}_{\! \scriptscriptstyle T}},
\ee
and, hence, we have
\bea
\left \langle 
\hat{\Phi}_{n}(x_1) \hat{\Phi}_{n}(x_2) 
\hat{\Phi}_{n}(x_3) \hat{\Phi}_{n}(x_4)
\right \rangle_{ \mbox{}_{\! \scriptscriptstyle T}} \!
&=&
\left \langle 
\hat{\Phi}_{n}(x_1) \hat{\Phi}_{n}(x_2) 
\right \rangle_{ \mbox{}_{\! \scriptscriptstyle T}}\!
\left \langle 
\hat{\Phi}_{n}(x_3) \hat{\Phi}_{n}(x_4)
\right \rangle_{ \mbox{}_{\! \scriptscriptstyle T}}
+\left \langle 
\hat{\Phi}_{n}(x_1) \hat{\Phi}_{n}(x_3) 
\right \rangle_{ \mbox{}_{\! \scriptscriptstyle T}}\!
\left \langle 
\hat{\Phi}_{n}(x_2) \hat{\Phi}_{n}(x_4)
\right \rangle_{ \mbox{}_{\! \scriptscriptstyle T}}
\nn    \\
&& 
+\left \langle 
\hat{\Phi}_{n}(x_1) \hat{\Phi}_{n}(x_4) 
\right \rangle_{ \mbox{}_{\! \scriptscriptstyle T}}\!
\left \langle 
\hat{\Phi}_{n}(x_2) \hat{\Phi}_{n}(x_3)
\right \rangle_{ \mbox{}_{\! \scriptscriptstyle T}}.
\eea
A similar expression holds for the four-point function of 
time-ordered products (consider  
$t_1 \!>\! t_2 \!>\! t_3 \!>\! t_4$
in the last equation).
These results can be easily generalized to thermal 
$2 N$-point functions
($N \!\in \! {\rm I\hspace{-0.4 ex}N}$). On the other hand, from
(\ref{thermal prop}), we can see that
$\langle \hat{a}_k^{(\a )} 
\rangle_{ \mbox{}_{ \scriptscriptstyle T}} \!=\! 0$ and,
following similar steps, we can show that the thermal 
$(2 N\!-\!1)$-point functions vanish.

%%%%%%%%%%%%%%%%%%%%%%%%%%%%%%%%%%%

\end{document}